\documentclass[%
reprint,
footinbib,
amsmath,amssymb,
aps,
prl,
]{revtex4-2}

\usepackage{graphicx}
\usepackage{dcolumn}
\usepackage{bm}

\usepackage{multirow}

\usepackage{amsmath,amssymb,amsfonts}
\usepackage{mathtools}
\usepackage{graphicx}
\usepackage{color}
\usepackage{appendix}
\usepackage{amsthm}
\usepackage{tikz}
\usepackage{youngtab}
\usepackage{mleftright}
\usepackage{url}
\usepackage[most]{tcolorbox}
\usepackage[utf8]{inputenc}

\usepackage{tikz}
\usetikzlibrary{shapes,snakes,positioning,automata,arrows.meta,calc,decorations.markings,math,arrows.meta}
\tikzstyle{vertex}=[circle, draw, inner sep=0pt, minimum size=6pt]
\usepackage{xcolor}
\usepackage{subfig}

\usepackage[export]{adjustbox}

\usepackage{algorithmic}

\usepackage{geometry}
\newcommand{\nn}{\nonumber}

\def\a{\alpha}
\def\b{\beta}

\def\d{\delta}

\def\t{\tau}

\def\w{\omega}
\def\<{\langle}
\def\>{\rangle}

\def\ha{{\hat{a}}}
\def\hb{{\hat{b}}}

\def\tu{{\tilde{0}}}

\def\t2{{\tilde{1}}}

\def\ha{{\hat{a}}}
\def\hb{{\hat{b}}}
\def\hc{{\hat{c}}}
\def\had{{\hat{a}^\dagger}}
\def\hbd{{\hat{b}^\dagger}}
\def\hcd{{\hat{c}^\dagger}}

\usepackage[normalem]{ulem}

\makeatletter
\newcommand{\newparallel}{\mathrel{\mathpalette\new@parallel\relax}}
\newcommand{\new@parallel}[2]{%
	\begingroup
	\sbox\z@{$#1T$}
	\resizebox{!}{\ht\z@}{\raisebox{\depth}{$\m@th#1/\mkern-5mu/$}}%
	\endgroup
}
\makeatother

\begin{document}

	\title{Exponentially Enhanced Scheme for the Heralded Qudit GHZ State\\in Linear Optics}

	\author{Seungbeom Chin}
	\email{seungbum.chin@oist.jp}
	\affiliation{Okinawa Institute of Science and Technology Graduate University, Okinawa 904-0495, Japan\\Department of Electrical and Computer Engineering, Sungkyunkwan University, Suwon 16419, Korea}

    \author{Junghee Ryu}
    \affiliation{Center for Quantum Information R$\&$D, Korea Institute of Science and Technology Information, Daejeon 34141, Korea\\
    Division of Quantum Information, KISTI School, Korea University of Science and Technology, Daejeon, 34141, Korea}

	\author{Yong-Su Kim}
	\affiliation{Center for Quantum Information, Korea Institute of Science and Technology (KIST), Seoul, 02792, Korea \\
		Division of Quantum Information Technology, KIST School, Korea University of Science and Technology, Seoul 02792, Korea}

	
	\begin{abstract}

High-dimensional multipartite entanglement plays a crucial role in quantum information science. However, existing schemes for generating such entanglement become complex and costly as the dimension of quantum units increases.  In this work, we overcome the limitation by proposing a significantly enhanced linear optical heralded scheme that generates the $d$-level $N$-partite GHZ state with single-photon sources and linear operations. Our scheme requires $dN$ photons, which is the minimal required photon number, with substantially improved success probability from previous schemes. It employs linear optical logic gates compatible with any qudit encoding system and can generate generalized GHZ states with installments of beamsplitters. With efficient generations of high-dimensional resource states, our work opens avenues for further exploration in high-dimensional quantum information processing.
	\end{abstract}
	

 \maketitle

\paragraph{Introduction.---} 
A qubit, which is comprised of a two-level quantum system and functions, is the quantum counterpart to the classical bit. It is generally considered the basic unit that stores and delivers information in quantum tasks. 
However, many natural systems have higher degrees of freedom, such as frequencies and paths of particles. Quantum information theory provides a new way of encoding such multi-level systems, which is called the \emph{qudit}. Qudits provide larger Hilbert spaces than qubits  for encoding and manipulating information.
Higher-dimensional quantum systems exhibit distinctive properties that are unattainable with qubit-based systems such as contextuality~\cite{budroni2022kochen}. They also provide a handy platform to simulate several quantum systems~\cite{tacchino2021proposal,chicco2023proof}. 


 Numerous advantages arise from entangling multiple quantum units in quantum information processing.
  The Bell state of qubits facilitates quantum enhancements over classical counterparts~\cite{shor1994algorithms, bennett1993teleporting}. Beyond two-qubit systems, quantum entanglement in multipartite and/or higher-dimensional systems is essential for new quantum technologies such as quantum computing~\cite{flamini2018photonic,wang201818,imany2019high,erhard2020advances,wang2020integrated},  quantum cryptography~\cite{groblacher2006experimental,mirhosseini2015high} quantum communication protocols~\cite{vaziri2002experimental,dada2011experimental,cozzolino2019high,langford2004measuring}.

On the other hand, the intricate structure of qudit multipartite entanglement brings about  several theoretical and practical challenges such as classification problems according the separability of states~\cite{walter2016multipartite} and the maximal quantum violation by non-maximally entangled qutrit bipartite states~\cite{collins2002bell, acin2002quantum}.
Another crucial problem lies on the \emph{generation of qudit mulipartite entangled states} for the practical performances of quantum tasks. The increased degrees of freedom in the system generally make the generation process significantly more complex and expensive.

We contribute to overcoming this limit by providing a \emph{significantly enhanced heralded scheme for generating the qudit multipartite entanglement in linear optical networks (LONs)}. 
The LON is one of promising systems that can feasibly generate quantum resource states. 
Non-destructive entanglement generations by heralding are possible in LONs by utilizing ancillary particles to herald the successful operation events without direct detections of the target states. Therefore,  photons entangled by heralding become proper resources for quantum computations with multiple gates~\cite{gimeno2015three,gubarev2020improved} and also facilitates loophole-free Bell tests~\cite{zhao2019higher,weston2018heralded}.
However, the number of additional particles and modes that the heralded schemes require increases as the dimension of entanglement becomes higher, which makes the circuit structures much more intricate. Therefore, it is prerequisite to design optimized circuits that achieve the target state with minimal resources for the efficient high-dimensional quantum information processing.

In this work, we propose a heralded scheme to generate one of essential qudit mutipartite entangled states, i.e., the $N$-partite $d$-level GHZ state $|GHZ_N^d\>$:
\begin{align}
  |GHZ_N^d\> = \frac{1}{\sqrt{d}}\sum_{j=0}^{d-1}|j\>_1\otimes \cdots \otimes |j\>_N 
\end{align} where $|j\>_k$ denotes the $k$th qudit being in the $j$th state.
The qudit GHZ state reveals intriguing non-classical properties of $N$-partite $d$-level systems such as the GHZ paradox~\cite{cerf2002greenberger,lee2006greenberger,tang2013greenberger, ryu2013greenberger} and Bell inequalities~\cite{son2006generic}. It also becomes a resource state to generate qudit cluster states for the high-dimensional measurement-based quantum computing (MBQC)~\cite{paesani2021scheme,yamazaki2024linear}.


Schemes for the qudit GHZ state in various physical setups have been proposed in, e.g., Refs.~\cite{erhard2018experimental,lee2022entangling,bhatti2023generating,paesani2021scheme, bell2022protocol,xing2023preparation, bhatti2024heralding} (see Appendix C for a comparison of these schemes). 
Particularly, Ref.~\cite{paesani2021scheme} first suggested a linear optical heralded scheme for $|GHZ_N^d\>$ based on the zero-transmission laws (ZTL) in quantum Fourier transforms (QFTs). However, it suffers from low efficiency and exponentially increasing photon numbers as $N$ and $d$ grows.

In this work, we  propose a scheme that generates the GHZ state of any $N$ and $d$ with the most feasible optical elements, i.e., single photons and linear optical operators as in Ref.~\cite{paesani2021scheme}.
Our scheme has several manifest advantages over formerly suggested schemes: First, it is an efficient circuit with \emph{$dN$ photons, which is the minimal photon number for generating $|GHZ_N^d\>$} (see Appendix D for the proof) and \emph{exponentially higher success probability than that in Ref.~\cite{paesani2021scheme}.}
Second, it can easily control the probability amplitudes of each term in the GHZ state by simply placing beamsplitters on the paths of photons, which result in \emph{the generation of the generalized qudit GHZ state}. Third, it consists of \emph{linear optical logic gates that any type of qudit encoding (e.g., optical angular momentum (OAM), multi-rail, time-bin, etc.) can establish.}  

\begin{figure}
    \centering
    \includegraphics[width=.45\textwidth]{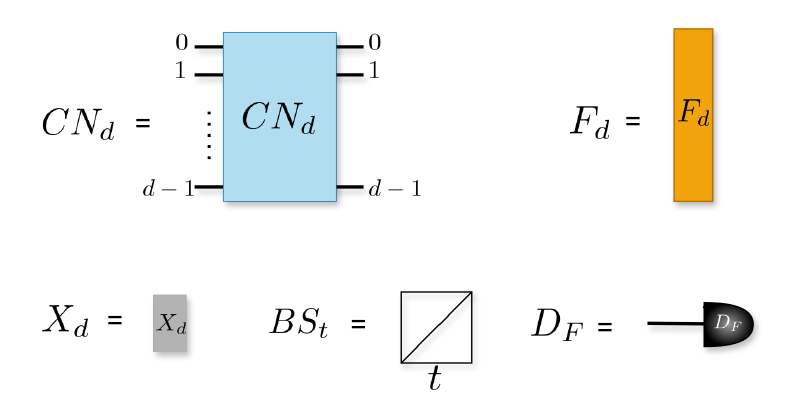}
    \caption{The list of linear optical logic gates that will be used to design the qudit GHZ generating circuits.}
    \label{fig:gates}
\end{figure}

\paragraph{Linear optical logic gates for generating $|GHZ_N^d\>$.---}
We first introduce a set of linear optical logic gates, which will compose our scheme. The list of the gates with their pictorial notations is given in Fig.~\ref{fig:gates}.

$ $\\
1.~$CN_d$: The $d$-level generalized CNOT gate between the spatial mode $j$ $(\in \{0,1,\cdots, d-1 \})$ and the internal state $s$ $(\in \{0,1,\cdots, d-1\})$, which works on states as
    \begin{align}\label{cnd}
      &|s\>_j = \ha^\dagger_{j,s}|vac\>\nn \\
      \to&~
      CN_{d}|s\>_{j} = 
      |s\>_{j\oplus_d s} = \ha^\dagger_{j\oplus_d s, s}|vac\>
      \end{align} ($\had_{j,s}$ is a photon creation operator, $|vac\>$ is the vacuum state, and  $\oplus_d$ is the addition modular $d$).
Its inverse $\overline{CN_d}$ works as
    \begin{align}\label{cndbar}
      &|s\>_j = \ha^\dagger_{j,s}|vac\>\nn \\
      \to
      &~\overline{CN}_d|s\>_j =  |s\>_{j\ominus_d s} = \ha^\dagger_{j\ominus_ds, s}|vac\>
     \end{align}  where $\ominus_d$ is the subtraction modular $d$. 
    These operators split states in the same spatial mode with different internal states into different spatial modes (see Appendix A for a more detailed explanation).  

    For $d=2$, we can build these operators with polarized photons. 
    By setting $\{H,V\} = \{0,1\}$ ($H$=horizontal, $V$=vertical), $CN_2$ and $\overline{CN}_2$ are both implemented with a polarizing beam splitter (PBS).\\
2. $F_{d}$: The internal state Fourier-transformation gate,
   \begin{align}
     F_d|s\>_j=|\tilde{s}\>_j= \frac{1}{\sqrt{d}}\sum_{r=0}^{d-1}\omega^{sr}|r\>_j~~~(\omega = e^{i\frac{2\pi}{d}}).  
    \end{align} 
 $F_d$ changes the computational basis of the internal state into its Fourier-transformed basis. For $d=2$, $F_2$ can be installed in the polarization encoding with a half-wave plate (HWP). \\
3.~$X_d$: The $d$-level unitary generalization of $\sigma_x$ in qubit,
\begin{align}
 X_d=\sum_{k=0}^{d-1}|k\oplus 1\>\<k|,   
\end{align} under which $|k\>$ transforms to $|k\oplus 1\>$.\\
4.~$BS_t$: A beam splitter that transmits a photon with transmissivity $t$.\\
5. $D_F$: Number resolving detections in the Fourier-transformed basis, i.e., measurements in $\{|\tilde{s}\>\}_{s=0}^{d-1}$ and postselections of the cases when only one particle arrives at the detector. This is achieved by combining $F_d$ and $CN_d$, i.e.,
\begin{align}
    \includegraphics[width=.3\textwidth]{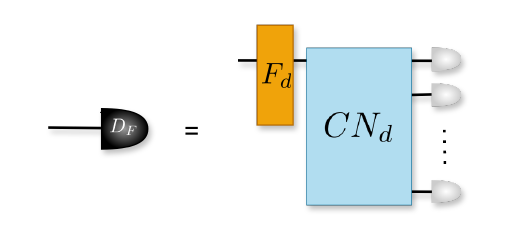} \nn 
\end{align} where $
\begin{gathered}\includegraphics[width=1cm]{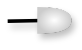}\end{gathered}$ denotes the particle-number-resolving detector.



\begin{center}
\begin{figure}[t]  \includegraphics[width=.5\textwidth]{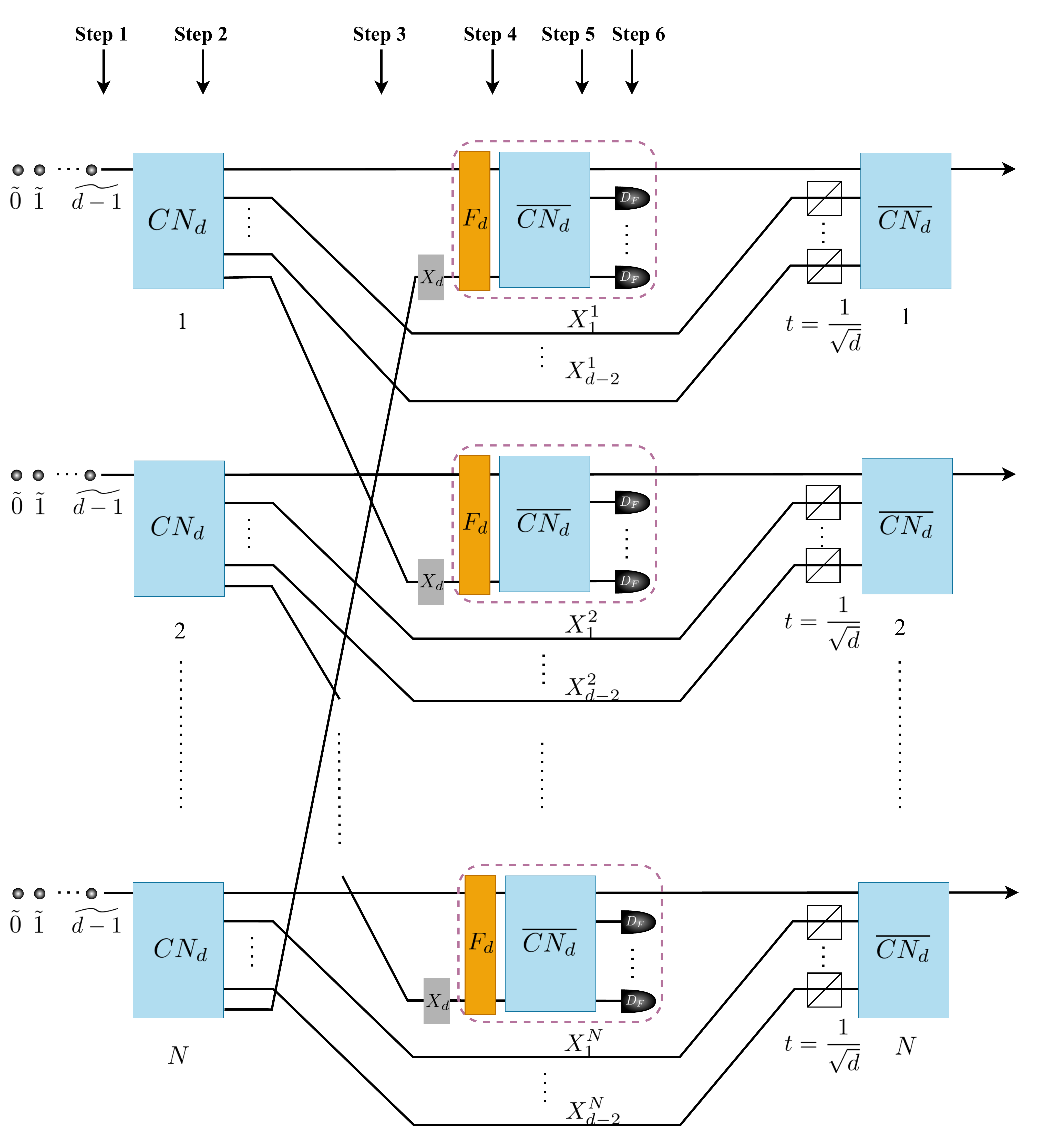}
    \caption{The linear optical circuit that generates $|GHZ_N^d\>.$ Beam splitters are attached at the end of the $d$ sub-modes of the first spatial subsystem $1$ to control the probability amplitudes of the terms in the target state, hence we can obtain the generalized qudit GHZ states.}
    \label{fig:graph_to_circuit}
\end{figure}
\end{center}

\paragraph{Linear optical circuits for generating $|GHZ_N^d\>$.---}
By combining the gates listed in Fig.~\ref{fig:gates}, we can design a circuit that generates $|GHZ_N^d\>$ for any $N$ and $d$.
Our scheme is given in Fig.~\ref{fig:graph_to_circuit}, which can be considered a linear optical implementation for the sculpting scheme in Ref.~\cite{chin2024shortcut} to generate the same target state. 
While the sculpting scheme in Ref.~\cite{chin2024shortcut} is a conceptual one with the assumption of the existence of high-dimensional deterministic annihilation operators, here we demonstrate that linear optical logic gates can combine to probabilistically generate the qudit GHZ state with heralding. Another crucial advantage of our scheme over Ref.~\cite{chin2024shortcut} is that the generalized qudit GHZ state can be easily generated by controlling the transmissivity of $BS_t$ in Fig.~\ref{fig:graph_to_circuit}.
See Appendix B for a more detailed explanation on these points.

In each input mode of the circuit in Fig.~\ref{fig:graph_to_circuit}, we inject $d$ photons with different internal states in the same mode. The operation of each dashed box in the circuit corresponds to the boson subtraction of $(d-1)$ particles, hence $(d-1)N$ subtraction of photons in total.
$BS_t$s are attached to control the probability amplitude of the target state, hence we can obtain different states in the class of generalized GHZ states by changing the transmissivity of the $BS_t$s.
It is worth noting that \emph{our circuit inherits the cyclic symmetry of the qudit GHZ state.}

 The success probability $P_{suc}(N,d)$ for generating $|GHZ_N^d\>$ is given by
\begin{align}
    P_{suc}(N,d) = d\Big(\frac{d!}{d^d} \Big)^{2N}. 
\end{align} 
When $d=2$, by considering $CN_2$ and $\overline{CN}_2$ with PBS and $F_d$ with HWP,  this circuit corresponds to the qubit GHZ generation scheme proposed in Ref.~\cite{chin2023graphs}. 
Appendix B demonstrates how the circuit in Fig.~\ref{fig:graph_to_circuit} generates $|GHZ_N^d\>$ by heralding.

\begin{center}
\begin{figure}[t]  \includegraphics[width=.5\textwidth]{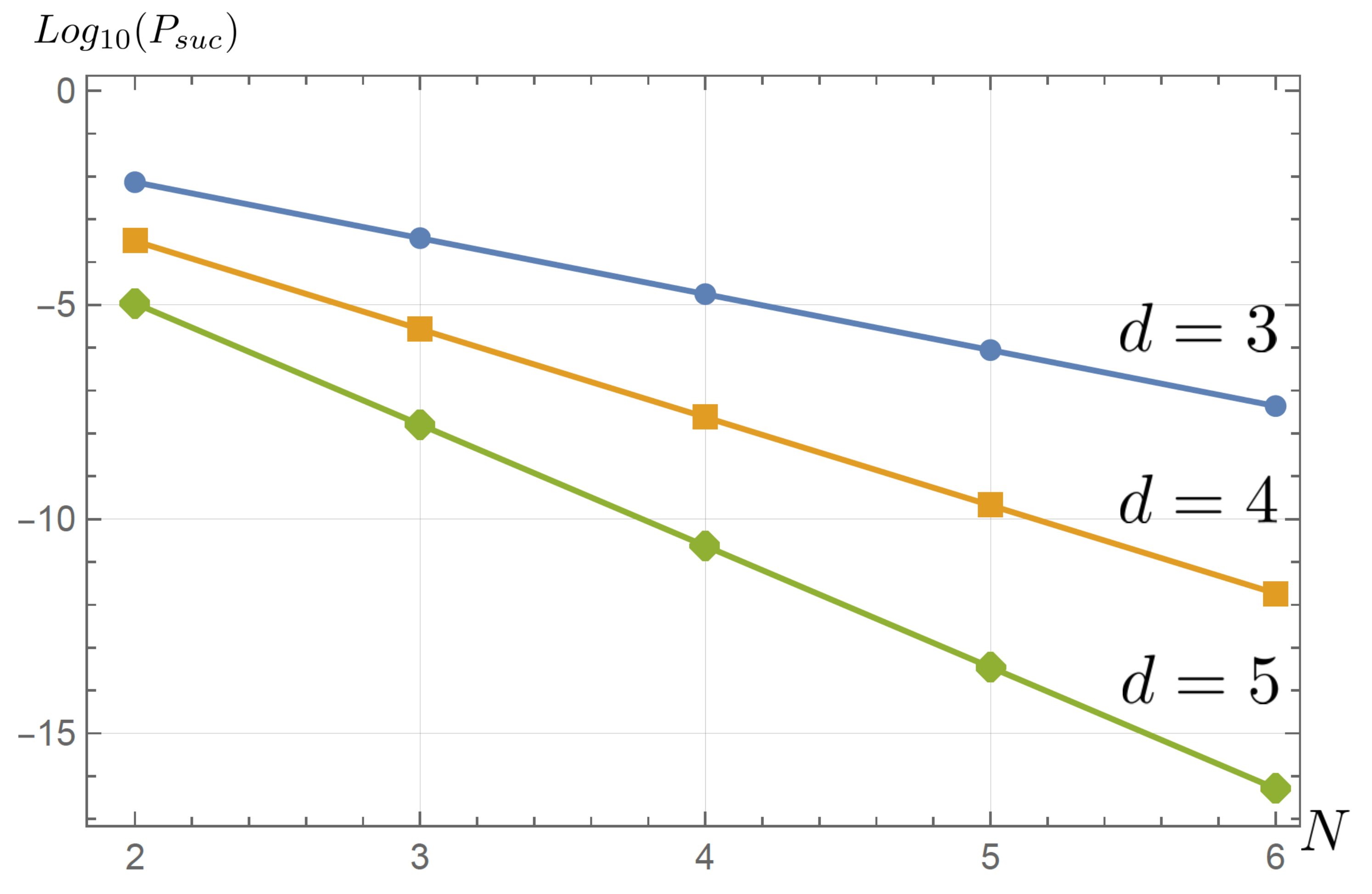}
    \caption{The logarithmic values of $P_{suc}(N,d)$ for $d=3,4,5$ as $N$ increases from 2 to 6. We can see that $P_{suc}(2,3)$ and $P_{suc}(3,3)$ have the order of $10^{-2}$  and $10^{-4}$ respectively. Our scheme is substantially more efficient for $N\ge 3$ than Ref.~\cite{paesani2021scheme} both in the photon numbers and success probabilities, which is explained in detail in Appendix C.2.}
    \label{fig:psuc}
\end{figure}
\end{center}

Fig.~\ref{fig:psuc} enumerates the numerical values of $Log_{10}(P_{suc}(N,d))$ for $d=3,4,5$. 
Compared to the schemes using the zero-transmission law (ZTL) in Fourier transform interferometers in Ref.~\cite{paesani2021scheme} (which we call `ZTL scheme' from now on), our scheme is substantially more efficient for $N\ge 3$ both in the photon numbers and success probabilities. Appendix C.2 presents a detailed comparative explanation on the efficiency of two schemes.

\begin{center}
       \begin{figure}[t]
           \includegraphics[width=.5\textwidth]{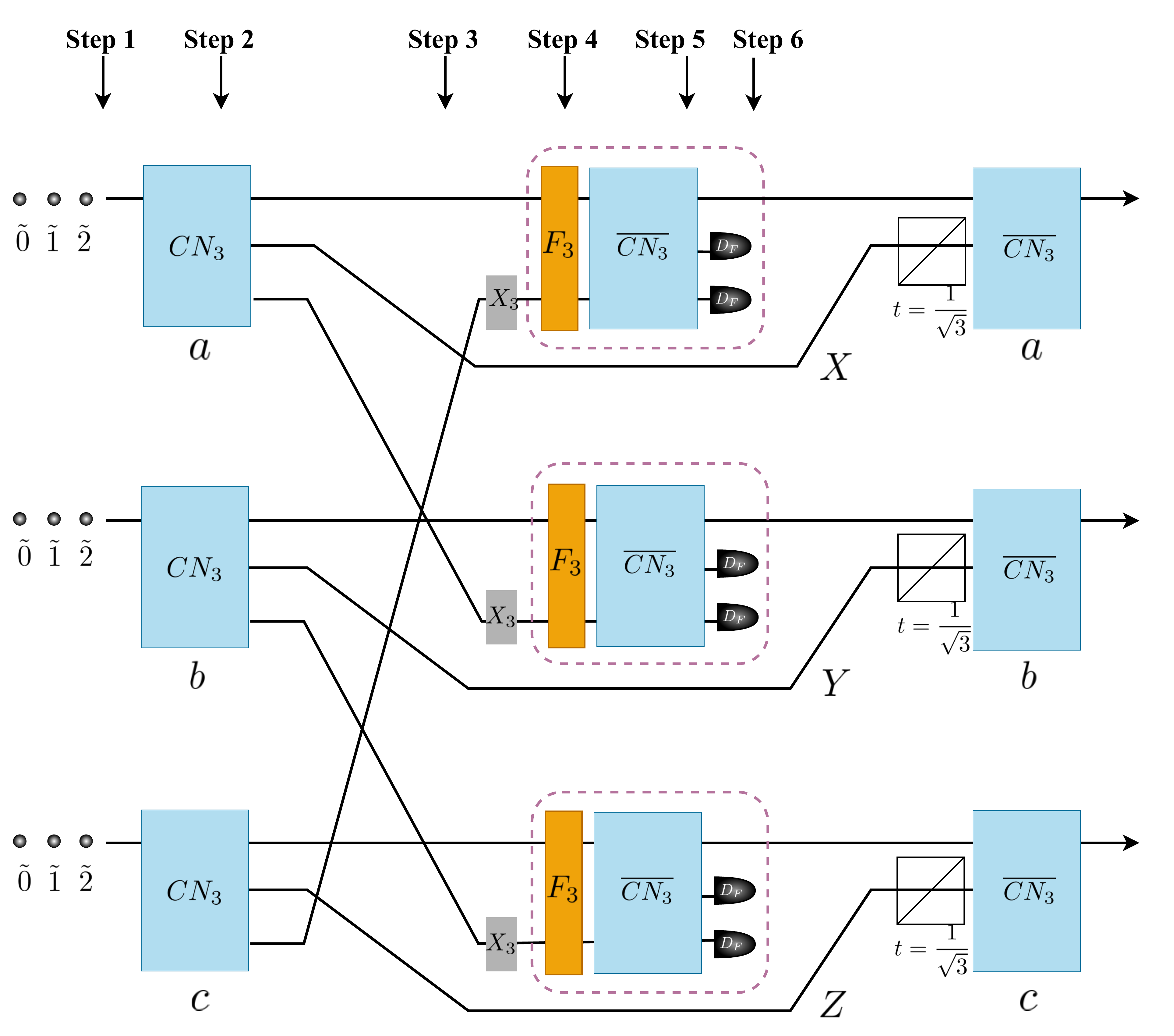}
           \caption{$|GHZ_3^3\>$ example of our scheme. We need a BS with transmissivity $\frac{1}{\sqrt{3}}$ on $X$, $Y$ and $Z$ to carve the extra amplitude of $|111\>$.}
           \label{fig:qutrit_GHZ}
       \end{figure}
\end{center}

\paragraph{$|GHZ_3^3\>$ example.---} As proof of concept, we provide here a step-by-step analysis on the $N=d=3$ example (Fig.~\ref{fig:qutrit_GHZ}). There are three spatial subsystems $\{a,b,c\}$ that has three sub-modes respectively, hence the photon creation operators are denoted as $\{\had_{j,s}, \hbd_{k,r}, \hcd_{l,t} \}$ where $\{j,k,l\}$ denotes the spatial sub-modes and $\{s,r,t\}$ the qudit states ($j,k,l,s,r,t \in \{0,1,2\}$). 

$ $\\
Step 1.~Initial state preparation: We inject three photons of orthogonal internal states with each other into the 0th modes of $a$, $b$, and $c$ respectively, i.e., 
    \begin{align}
&\ha^\dagger_{0,0}\ha^\dagger_{0,1}\ha^\dagger_{0,2}\hb^\dagger_{0,0}\hb^\dagger_{0,1}\hb^\dagger_{0,2} \hat{c}^\dagger_{0,0}\hat{c}^\dagger_{0,1}\hat{c}^\dagger_{0,2}|vac\>,
        \end{align} of which we Fourier-transform the internal state basis:
     \begin{align}
&\ha^\dagger_{0,\tilde{0}}\ha^\dagger_{0,\tilde{1}}\ha^\dagger_{0,\tilde{2}}\hb^\dagger_{0,\tilde{0}}\hb^\dagger_{0,\tilde{1}}\hb^\dagger_{0,\tilde{2}} \hat{c}^\dagger_{0,\tilde{0}}\hat{c}^\dagger_{0,\tilde{1}}\hat{c}^\dagger_{0,\tilde{2}}|vac\> . 
        \end{align}       \\
Step 2.~$CN_3$ operation: This operator works on the basis of $\{|j\>\}_{j=0}^{2}$, hence the initial state is transformed as
    \begin{align}
 \frac{1}{\sqrt{3}^{9}} & \big(\ha^{\dagger 3}_{0,0} +\ha^{\dagger 3}_{1,1}+\ha^{\dagger 3}_{2,2} -3\ha^{\dagger}_{0,0}\ha^{\dagger}_{1,1}\ha^{\dagger}_{2,2}\big) \nn \\
 &\times
\big(\hb^{\dagger 3}_{0,0} +\hb^{\dagger 3}_{1,1}+\hb^{\dagger 3}_{2,2} -3\hb^{\dagger}_{0,0}\hb^{\dagger}_{1,1}\hb^{\dagger}_{2,2}\big)   \nn \\
& \times \big(\hat{c}^{\dagger 3}_{0,0} +\hat{c}^{\dagger 3}_{1,1}+\hat{c}^{\dagger 3}_{2,2} -3\hat{c}^{\dagger}_{0,0}\hat{c}^{\dagger}_{1,1}\hat{c}^{\dagger}_{2,2}\big)|vac\> 
    \end{align} \\
Step 3. Rewiring: Now we rewire so that creation operators $\{\had_{1,1}, \hbd_{1,1}, \hcd_{1,1}\}$ are sent to $\{\hat{X}^\dagger_{1}, 
 \hat{Y}^\dagger_{1}, \hat{Z}^\dagger_{1} \}$ and $\{\had_{2,2}, \hbd_{2,2}, \hcd_{2,2}\}$ to $\{\hbd_{2,2}, \hcd_{2,2}, \had_{2,2}\}$ :
  \begin{align}
 \frac{1}{\sqrt{3}^{9}} & \big(\ha^{\dagger 3}_{0,0} +\hat{X}^{\dagger 3}_{1}+\hb^{\dagger 3}_{2,2} -3\ha^{\dagger}_{0,0}\hat{X}^{\dagger}_{1}\hb^{\dagger}_{2,2}\big) \nn \\
&\times  \big(\hb^{\dagger 3}_{0,0} +\hat{Y}^{\dagger 3}_{1}+\hat{c}^{\dagger 3}_{2,2} -3\hb^{\dagger}_{0,0}\hat{Z}^{\dagger}_{1}\hat{c}^{\dagger}_{2,2}\big)   \nn \\
&\times         \big(\hat{c}^{\dagger 3}_{0,0} +\hat{Z}^{\dagger 3}_{1}+\hat{a}^{\dagger 3}_{2,2} -3\hat{c}^{\dagger}_{0,0}\hat{X}^{\dagger}_{1}\hat{a}^{\dagger}_{2,2}\big)|vac\> 
    \end{align}
Step 4. $X_d$ and $F_3$: $F_3$ is applied to both wires in each subsystem, hence the state is transformed to
    \begin{align}
\frac{1}{\sqrt{3}^{9}} & \big(\ha^{\dagger 3}_{0,\tilde{0}} +\hat{X}^{\dagger 3}_{1}+\hb^{\dagger 3}_{2,\tilde{0}} -3\ha^{\dagger}_{0,\tilde{0}}\hat{X}^{\dagger}_{1}\hb^{\dagger}_{2,\tilde{0}}\big) \nn \\
\times & \big(\hb^{\dagger 3}_{0,\tilde{0}} +\hat{Y}^{\dagger 3}_{1}+\hc^{\dagger 3}_{2,\tilde{0}} -3\hb^{\dagger}_{0,\tilde{0}}\hat{Y}^{\dagger}_{1}\hc^{\dagger}_{2,\tilde{0}}\big) \nn \\
\times & \big(\hat{c}^{\dagger 3}_{0,\tilde{0}} +\hat{Z}^{\dagger 3}_{1}+\ha^{\dagger 3}_{2,\tilde{0}} -3\hat{c}^{\dagger}_{0,\tilde{0}}\hat{Z}^{\dagger}_{1}\ha^{\dagger}_{2,\tilde{0}}\big)       |vac\>  
        \end{align} 
Step 5. $\overline{CN_3}$      
    \begin{align}
&\frac{1}{3^9}
\Big[(\had_{0,0} +\had_{2,1} +\had_{1,2})^3  +\hat{X}^{\dagger 3}_{1}+ (\hbd_{2,0} +\hbd_{1,1} +\hbd_{0,2})^3  \nn \\
&~~~-3\sqrt{3}(\had_{0,0} +\had_{2,1} +\had_{1,2}) \hat{X}^{\dagger}_{1}(\hbd_{2,0} +\hbd_{1,1} +\hbd_{0,2}) 
\Big] \nn \\
&~\times \Big[(\hbd_{0,0} +\hbd_{2,1} +\hbd_{1,2})^3  + \hat{Y}^{\dagger 3}_{1} +(\hcd_{2,0} +\hcd_{1,1} +\hcd_{0,2})^3\nn \\
&~~~-3\sqrt{3}(\hbd_{0,0} +\hbd_{2,1} +\hbd_{1,2}) \hat{Y}^{\dagger}_{1}(\hcd_{2,0} +\hcd_{1,1} +\hcd_{0,2}) \Big]  \nn \\
&~\times 
\Big[(\hcd_{0,0} +\hcd_{2,1} +\hcd_{1,2})^3  +\hat{Z}^{\dagger 3}_{1}  + (\had_{2,0} +\had_{1,1} +\had_{0,2})^3\nn \\
&~~~-3\sqrt{3}(\hcd_{0,0} +\hcd_{2,1} +\hcd_{1,2}) \hat{Z}^{\dagger}_{1}(\had_{2,0} +\had_{1,1} +\had_{0,2}) 
\Big]|vac\> 
        \end{align}
Step 6. Postselections and feed-forward:
After the postselection of the cases when each detector observes only one photon, the terms that contribute to the final states are
\begin{align}
    \frac{1}{3^9}\Big[& 6^3 \had_{00}\had_{21}\had_{12} \hbd_{00}\hbd_{21}\hbd_{12} \hcd_{00}\hcd_{21}\hcd_{12} \nn \\ &+6^3\had_{20}\had_{11}\had_{02}\hb^\dagger_{20}\hb^\dagger_{11}\hb^\dagger_{02} \hcd_{20}\hcd_{11} \hcd_{02} \nn \\
&-(3\sqrt{3})^3 \hat{X}^\dagger_{1}\hat{Y}^\dagger_{1}\hat{Z}^\dagger_{1}
(\had_{1,1}\had_{2,1}+\had_{1,2}\had_{2,0})   \nn \\
&~~~\times (\hbd_{1,1}\hbd_{2,1}+ \hbd_{1,2}\hbd_{2,0})(\hcd_{1,1}\hcd_{2,1}+ \hcd_{1,2}\hcd_{2,0})\Big]|vac\>
\end{align}
After the postselection with $D_F$s and feedforwards, the final state that passes through $BS_t$ becomes
\begin{align}
    \frac{6^3}{3^9}\Big[|0,0,0\> - 3\sqrt{3}t^3 |1,1,1\> +|2,2,2\> \Big].
\end{align}
Setting the transmissivity $t=\frac{1}{\sqrt{3}}$, the final state becomes the tripartite qutrit GHZ state.
The success probability is given by
\begin{align}
    P_{suc}(3,3) = 3\times (\frac{2}{9})^{6} \sim 3.6\times 10^{-4}.
\end{align}


\paragraph{Construction of linear optical schemes in OAM  and multi-rail encodings.---}
The linear optical logic gates in Fig.~\ref{fig:gates} can be implemented with any qudit encoding system of linear optics.
Among them, we discuss the OAM and multi-rail encodings.  

In the OAM encoding, it is straightforward to establish the scheme, because the OAM beam splitters~\cite{zou2005scheme}, OAM-only Fourier transformation operators~\cite{kysela2020fourier}, and OAM high-dimensional $X$ gate~\cite{vashukevich2022high} implement the $CN_d$, $F_d$, and $X_d$ gates respectively. 

 In the multi-rail encoding, a permutation of photon paths becomes the $CN_d$ gate and the $d$-partite port that Fourier transforms the spatial states of photons becomes the $F_d$ gate (see Appendix E).
An advantage of the multirail encoding circuit is that $CN_d$ and $\overline{CN}_d$ is achieved with the permutation of paths, hence \emph{the final circuit consists of simpler optical elements}, i.e., $d$-partite ports including BSs, permutations of paths, and photon detectors.  

\paragraph{Discussions.---}
Our scheme has several manifest advantages in using the qudit GHZ state as a resource for high-dimensional quantum information processing. 
First,  as we have already demonstrated, our scheme enables more efficient generations of the resource states with the minimal number of single-photons.
Given that 20 single photons can be successfully generated and controlled in linear optical systems~\cite{wang2019boson}, we expect relatively low-dimensional qudit GHZ states to be generated sooner or later using in laboratories or Quandela platform~\cite{maring2024versatile}.
Second, our scheme can generate generalized qudit GHZ states, which is adequate to verify the optimal state for the violation of Bell-CHSH nonlocality test~\footnote{While the maximally entangled state entails the maximal Bell-CHSH violation in $d=2$ systems, the scenario becomes more complicated for higher dimensional systems. Ref.~\cite{collins2002bell,acin2002quantum} shows that the maximal Bell-CHSH violation occurs for a non-maximally entangled state, which is also verified numerically for the $N=d=3$ case in Ref.~\cite{laskowski2014noise}. Therefore, for the experiments of those maximal violations, schemes for generalized qudit GHZ states with relatively asymmetric amplitudes are indispensable. Our scheme easily achieve the condition by varying the transmissivity of beamspliters.}.
Third, our scheme consists of linear optical logic gates that any type of qubit encoding (e.g., optical angular momentum (OAM), multi-rail, time-bin, etc.) can realize.

It is worth emphasizing that the methodology of composing linear optical logic gates based on the linear quantum graph picture (LQG picture)~\cite{chin2021graph,chin2024shortcut} (Appendix B.1 for a brief explanation) can be useful to find more general types of qudit multipartite entanglement. For example, Ref.~\cite{chin2024creating} proposed heralded schemes for $N$-partite $N$-level symmetric and anti-symmetric states based on the method. We expect to construct more rigorous and general formalism of the LQG picture that will provide a more systematic path for the heralded generations of essential qudit entanglement such as absolutely maximally entangled states.

In addition, it will be an intriguing problem to find the upper bound of the success probability for generating $|GHZ_N^d\>$ with minimal photon number, $dN$. We expect that the mathematical technique to determine the required minimal photon number (Appendix D) could be exploited to search for the solution.

\paragraph{Acknowledgements.---} SC is grateful to Prof. William J. Munro and  Prof. Jung-Hoon Chun for their support on this research. SC is supported by
National Research Foundation of Korea (NRF, 2019R1I1A1A01059964, RS-2023-00245747). JR acknowledges support from the National Research Foundation (NRF) of Korea grant funded by the Korea Government (Grant No. NRF-2022M3K2A1083890). YSK acknowledges support from the KIST institutional program (2E32941). 

\bibliographystyle{apsrev4-2}
\bibliography{qudit_GHZ}

\newpage	
\onecolumngrid	
\appendix

\section{Operations of $CN_d$, the $d$-level CNOT gate between the spatial mode and the internal state}\label{cnot_d}

Unlike the conventional generalized CNOT gate that works on two different qudits, 
$CN_d$ and $\overline{CN}_d$ here transform the spatial mode of a given photon according to its internal state. They transform the two variables in the same way as the generalized CNOT gate between two qudits. The spatial mode and internal state play the role of the target and control states. 

More explicitly, the operation of $CN_d$ given by Eq.~(2) of the main content can be enumerated as the following table: 
\begin{table}[h]
	\begin{tabular}{|l|l|l|l|}
		\hline
		$|0\>_0\to |0\>_0$  & $|1\>_0\to |1\>_1$   & $\cdots$ &  $|d-1\>_0\to |d-1\>_{d-1}$     \\
		\hline
		$|0\>_1\to |0\>_1$ & $|1\>_1\to |1\>_2$   & $\cdots$ &  $|d-1\>_1\to |d-1\>_{0}$   \\
		\hline
		$\vdots$ &  $\vdots$   & $\vdots$ &  $\vdots$  \\
		\hline 
		$|0\>_{d-2}\to |0\>_{d-2}$ & $|1\>_{d-2}\to |1\>_{d-1}$ & $\cdots$ &  $|d-1\>_{d-2}\to |d-1\>_{d-3}$ \\
		\hline
		$|0\>_{d-1}\to |0\>_{d-1}$ & $|1\>_{d-1}\to |1\>_{0}$ & $\cdots$ &  $|d-1\>_{d-1}\to |d-1\>_{d-2}$  \\ 
		\hline 
	\end{tabular}
\end{table}

Comparing the transformations in the same row of the table, we can see that that $CN_d$ splits photons of different internal states in the same spatial mode into different spatial modes.

\section{Comprehensive analysis on the qudit $N$-partite GHZ state generation scheme}\label{N_GHZ}

Here we provide a comprehensive explanation on the linear optical circuit for generating the qudit GHZ state given in Figure~2 of the main content. We first explain how we designed our scheme motivated by the graph picture of linear quantum networks (or linear quantum graph picture, LQG picture)~\cite{chin2021graph,chin2024shortcut} and then show that the scheme indeed generates the expected target state.

\subsection{From sculpting operators to photon subtraction operators}\label{sculpting_bigraph}

The LQG picture provides a set of correspondence relations between graphs and many-particle systems that linearly evolve. This graph picture was introduced to provide  systematical methods to search for schemes that can generate genuine multipartite entanglement. Ref.~\cite{chin2021graph} first introduced the LQG picture and found several postselected schemes for multipartite entangled states. Ref.~\cite{chin2024shortcut} generalized the formalism and applied it to find schemes for generating multipartite entanglement with the sculpting scheme~\cite{karczewski2019sculpting} (which generates an N -partite entangled state by
applying tailored single-boson annihilation operators). For the qubit case, the graph solutions for generating entanglement with sculpting schemes in Ref.~\cite{chin2024shortcut} can be implemented in linear optical systems probabilistically by a set of direct translation rule~\cite{chin2023graphs}. However, the sculpting scheme for generating high-dimensional GHZ state in Ref.~\cite{chin2024shortcut} has remained conceptual without any suggestion to experimentally execute the high-dimensional sculpting process in many physical system. 
And this process for qudits is considerably more complicated than the qubit case in which the particle paths are simply split using polarized beam splitters (PBS) by the Hong-Ou-Mandel effect. 

In this section, we explain the sculpting scheme for $|GHZ_N^d\>$ proposed in Ref.~\cite{chin2024shortcut} and 
how to construct linear optical logic gates that execute the high-dimensional boson annihilation operations for generating the qudit GHZ state.


In Ref.~\cite{chin2024shortcut}, a boson sculpting scheme for $|GHZ_N^d\>$ is introduced as follows:
\begin{enumerate}
	\item Prepare for $dN$ bosons in  $N$ spatial modes so that each boson states are orthogonal to each other, e.g.,
	\begin{align}\label{initial}
		|Init_{N,d}\> \equiv \prod_{j=1}^N(\had_{j,0}\had_{j,1}\cdots \had_{j,d-1})|vac\>,
	\end{align} where $\had_{j,s}$ denotes a boson in the $j$th subsystem ($j \in \{1,2,\cdots, N\}$) with internal qudit state $s$ $( \in \{0,1,\cdots, d-1\})$.
	\item Apply a boson sculpting operator
	\begin{align}\label{subtraction_operator}
		\hat{A}_{N,d}
		=\Big(\frac{1}{\sqrt{2}}\Big)^{(d-1)N}(\ha_{1,\tu}-\ha_{2,\widetilde{d-1}})^{d-1} (\ha_{2,\tu}-\ha_{3,\widetilde{d-1}})^{d-1} (\ha_{3,\tu}-\ha_{4,\widetilde{d-1}})^{d-1} \cdots  (\ha_{N,\tu}-\ha_{1,\widetilde{d-1}})^{d-1}.
	\end{align}	
	\item Using Identities 1 and 2 in Appendix  Section~\ref{demonstration}, the final state is given by the qudit GHZ state in the Fourier-transformed basis up to normalization:
	\begin{align}
		\Big(\frac{1}{\sqrt{2}}\Big)^{(d-1)N}\Big(|\tilde{0},\tilde{0},\cdots, \tilde{0}\> +|\tilde{1},\tilde{1},\cdots, \tilde{1}\>+\cdots +|\widetilde{d-1},\widetilde{d-1},\cdots, \widetilde{d-1}\>\Big) 
	\end{align}
\end{enumerate}
The graph representation of the operator~\eqref{subtraction_operator} is given by (see Table I of Ref.~\cite{chin2024shortcut})
\begin{align}\label{GHZ_graph}
	\includegraphics[width=4cm]{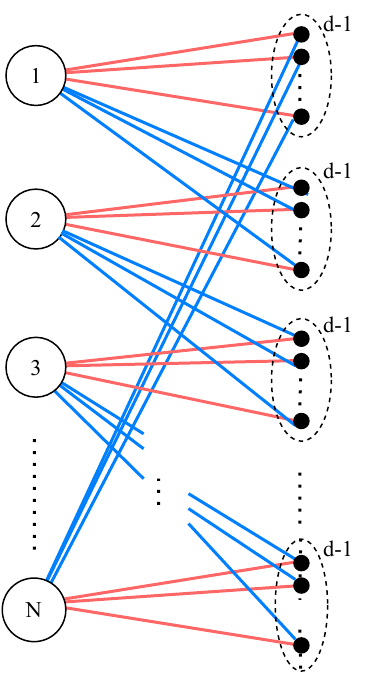} 
\end{align}
Decomposing the above graph, it consists of 
\begin{align}
	\begin{gathered}
		\includegraphics[width=4cm]{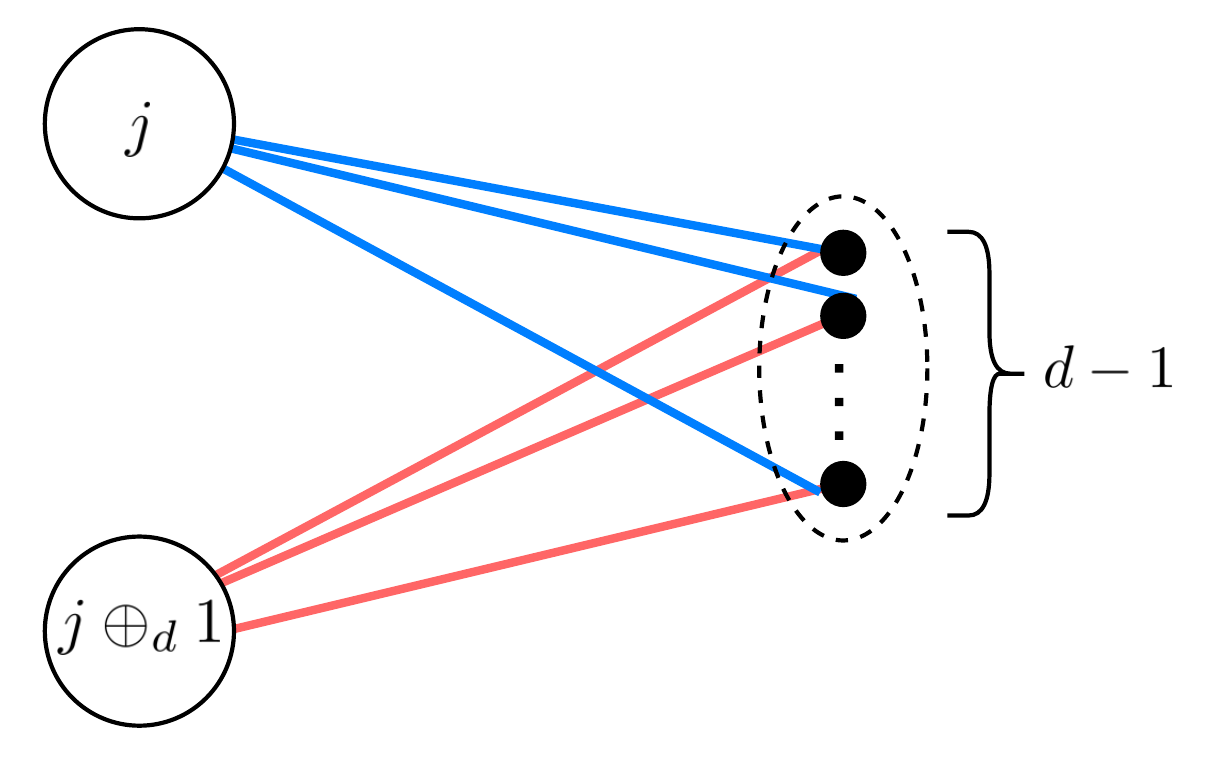}    
	\end{gathered},
	~~~~(j\in\{1,2,\cdots, N \},~\textrm{$\oplus_d$ is addition mod $d$}) 
\end{align}
which corresponds to the operator
\begin{align}
	\frac{1}{\sqrt{2}^{d-1}}(\ha_{j,\tilde{0}}-\ha_{j\oplus_d 1,\widetilde{d-1}})^{d-1}.
\end{align} Therefore, we need a linear optical element that subtracts $(d-1)$ photons, whose photon state is the superposition of $|j,\tilde{0}\>$ and $|j\oplus_d 1,\widetilde{d-1}\>$.

From the property of the $CN_d$ gate (Eq.~(2) of the main content), we see that photons of state $\tilde{0}$ ($\widetilde{d-1}$) comes from the 1st ($(d-1)$-th) mode of a $CN_d$ gate. And we detect $(d-1)$ photons in the Fourier-transformed bases to herald the subtraction operator.  By combining these observations, we can expect that the following correspondence holds from the LQG picture to the linear optical system:
\begin{align}
	\includegraphics[width=14cm]{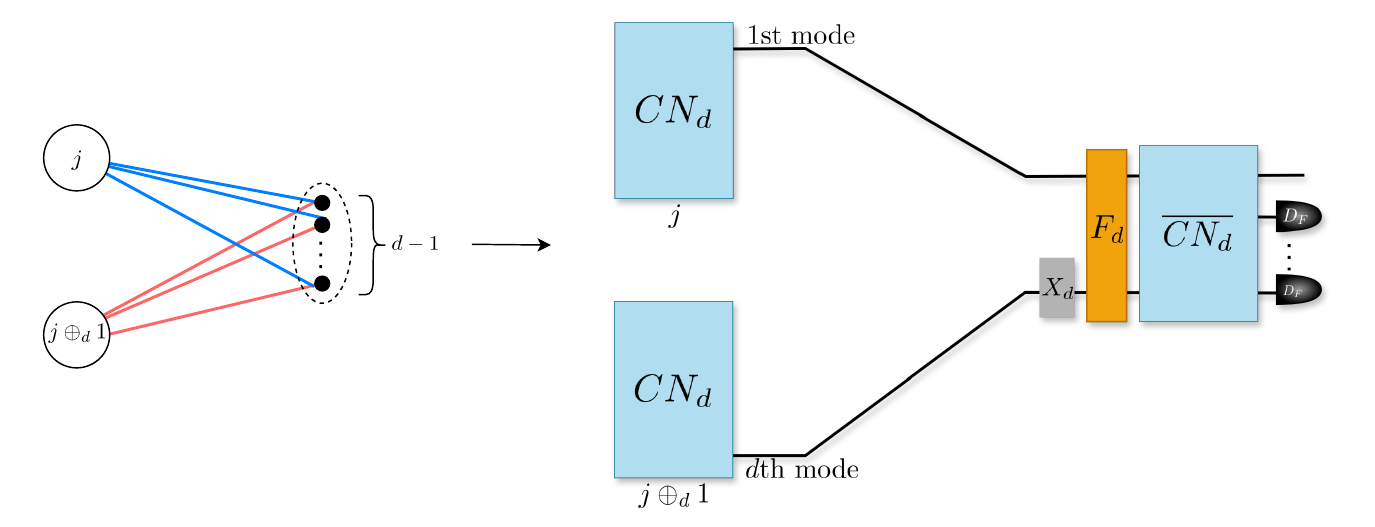} 
\end{align}
Note that $X_d$ is attached to the $d$-th mode of the $CN_d$ gate from the $(j\oplus_d 1)$-th mode to simplify the relative phases of the photons from the two different spatial modes. And photons from the $2$nd to the $(d-2)$-th modes of $CN_d$ gates must remain in the same subsystem.  

By connecting the $N$ sets of the above linear optical element based on the structure of the bigraph~\eqref{GHZ_graph}, we obtain the scheme in Figure~2  of the main content. 

\subsection{Demonstrations of the final state generated by the scheme in Figure~2 of the main content}\label{demonstration}

To demonstrate that the scheme in Figure~2 of the main content actually generates the qudit GHZ state, we need the following two identities that were first given in Ref.~\cite{chin2024shortcut} without proofs. Due to their significance in calculating our circuit operation, we restate them here with proofs.  

$ $\\
\noindent 
\textbf{Identity 1.} 
For $l\in \{ 0,1,\cdots ,d-1 \}$, 
\begin{align}
	(\ha_{\tu})^l(\ha_{\widetilde{d-1}})^{d-1-l}\prod_{s=0}^{d-1}\ha^\dagger_{s} |vac\>=(-1)^{d-1-l}\frac{l!(d-1-l)!}{\sqrt{d}^{d-2}}
	\ha^\dagger_{\widetilde{d-1-l}} |vac\>.
\end{align} Here all the operators are on the same subsystem, hence $\had_{s}$ and $\ha_{s}$ denote the creation and annihilation operators and  $s$ is the internal qudit state of the operators.  
\begin{proof}
	Using the following relations
	\begin{align}
		&\had_{\tilde{j}} = \frac{1}{\sqrt{d}}(\ha_{0}+\omega^{j}\ha_{1}+\cdots + \omega^{j(d-1)}\ha_{d-1}), \nn \\
		&\ha_{\tilde{j}} = \frac{1}{\sqrt{d}}(\ha_{0}+\omega^{-j}\ha_{1}+\cdots + \omega^{-j(d-1)}\ha_{d-1}),    
	\end{align} we have
	\begingroup
	\allowdisplaybreaks
	\begin{align}\label{identity_proof}
		&(\ha_{\tu})^l(\ha_{\widetilde{d-1}})^{d-1-l}\prod_{s=0}^{d-1}\ha^\dagger_{s} |vac\>\nn \\
		&=\frac{1}{\sqrt{d}^{d-1}}(\ha_{0}+\ha_{1}+\cdots + \ha_{d-1})^l (\ha_{0}+\omega\ha_{1}+\cdots + \omega^{(d-1)}\ha_{d-1})^{d-1-l} \prod_{s=0}^{d-1}\ha^\dagger_{s} |vac\> \nn \\
		&= 
		\frac{1}{\sqrt{d}^{d-1}}
		(\ha_{0}+\ha_{1}+\cdots + \ha_{d-1})^l 
		\sum_{\substack{j_1\neq j_2\neq \cdots\\ \neq j_{d-1-l} \\ =0} }^{d-1}\w^{(j_1+j_2+\cdots + j_{d-1-l})}\ha_{j_1}\ha_{j_2}\cdots \ha_{j_{d-1-l}}
		\prod_{s=0}^{d-1}\ha^\dagger_{s} |vac\> \nn \\
		&= 
		\frac{l!}{\sqrt{d}^{d-1}}
		\sum_{\substack{j_1\neq j_2\neq \cdots\\ \neq j_{d-1-l} \\ =0} }^{d-1}\w^{j_1+j_2+\cdots + j_{d-1-l}}(\sum_{q=0}^{d-1}\had_{q} - 
		\had_{j_1}- \had_{j_2} -\cdots - \had_{j_{d-1-l}})|vac\>
	\end{align}
	On the other hand, since
	\begin{align}
		\sum_{\substack{j_1\neq j_2\neq \cdots\\ \neq j_{d-1-l} \\ =0} }^{d-1}\w^{j_1+j_2+\cdots + j_{d-1-l}}
		=0
	\end{align} holds, 
	Eq.~\eqref{identity_proof} is simplified as
	\begin{align}
		&-\frac{l!}{\sqrt{d}^{d-1}}
		\sum_{\substack{j_1\neq j_2\neq \cdots\\ \neq j_{d-1-l} \\ =0} }^{d-1}\w^{j_1+j_2+\cdots + j_{d-1-l}}(
		\had_{j_1}+\had_{j_2 }+\cdots +\had_{j_{d-1-l}})|vac\>  \nn \\
		&= -\frac{l!}{\sqrt{d}^{d-1}}(d-1-l)
		\sum_{\substack{j_1\neq j_2\neq \cdots\\ \neq j_{d-1-l} \\ =0} }^{d-1} \w^{j_1+j_2+\cdots + j_{d-1-l}}
		\had_{j_1}|vac\> \nn \\
		&= (-1)^{d-1-l}\frac{l!(d-1-l)!}{\sqrt{d}^{d-1}}\sum_{j_1=0}^{d-1}\w^{(d-1-l)j_1}\had_{j_1}|vac\> \nn \\ 
		&= 
		(-1)^{d-1-l}\frac{l!(d-1-l)!}{\sqrt{d}^{d-2}}
		\ha^\dagger_{\widetilde{d-1-l}} |vac\>.
	\end{align}
	\endgroup
\end{proof} 
$ $\\
\textbf{Identity 2.} 
For $m \in \{1,\cdots, d-1 \}$,  
\begin{align}
	(\ha_{\tu})^m(\ha_{\widetilde{d-1}})^{d-m}\prod_{s=0}^{d-1}\ha^\dagger_{s} |vac\> = 0
\end{align}

\begin{proof}
	It is directly derived from Identity 1. 
\end{proof}



\begin{figure}[t]
	\centering 
	\includegraphics[width=.7\textwidth]{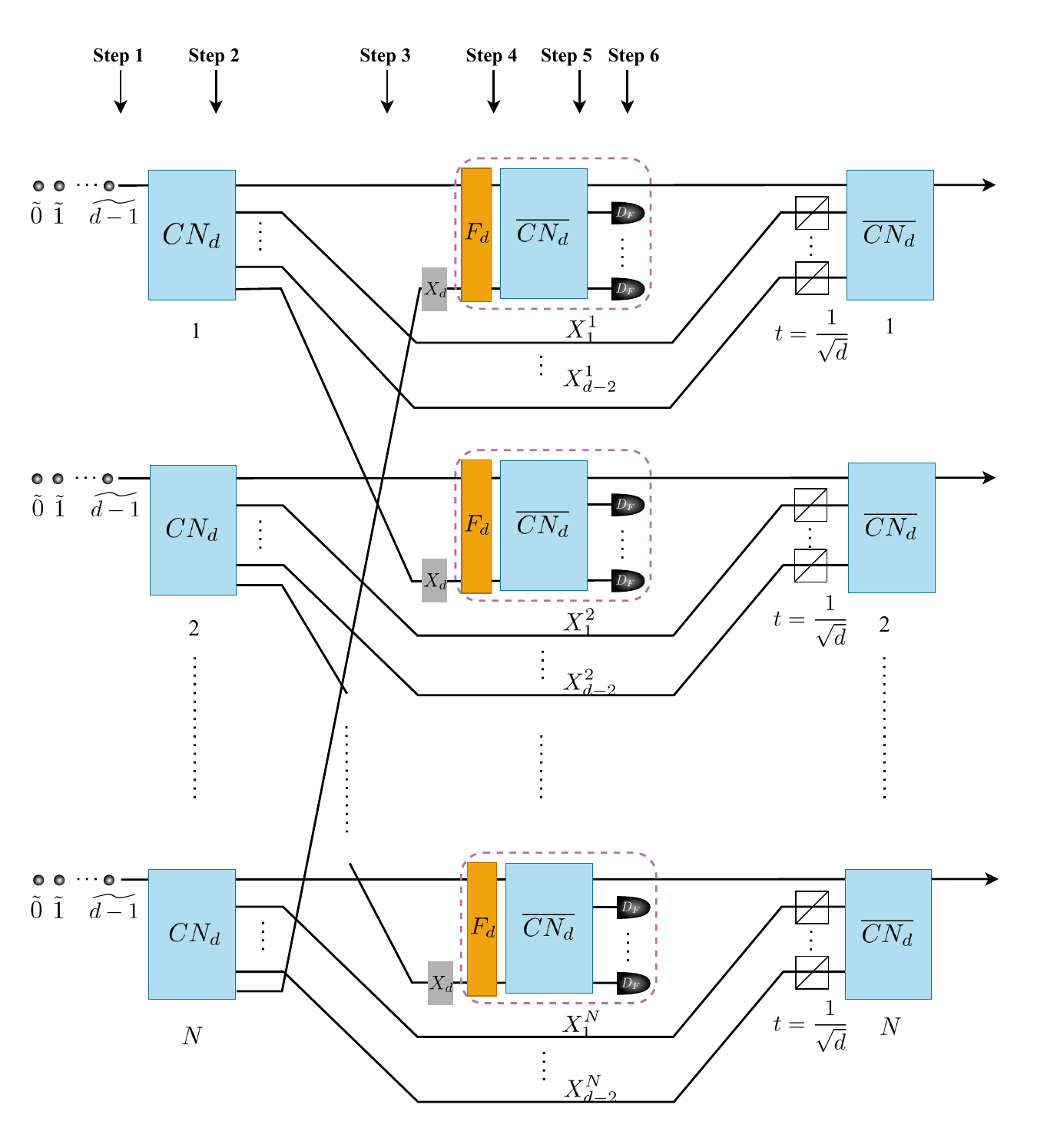}
	\caption{The linear optical circuit that generates $|GHZ_N^d\>.$ Beam splitters are attached at the end of the $d$ sub-modes of the first spatial subsystem $1$ to control the probability amplitudes of the terms in the target state, hence we can obtain the generalized qudit GHZ states.}
	\label{fig:graph_to_circuit_supp}
\end{figure}

There are $N$ spatial subsystems that has $d$ modes respectively in Figure~2 of the main content, hence the photon creation operator is denoted as $\ha^{j \dagger}_{k,s}$, which denotes that a photon is in the $j$th subsystem, $k$th mode, and $s$th internal state. 
Figure~\ref{fig:graph_to_circuit_supp} denotes the same optical circuit for the qudit GHZ state with Figure~2 of the main content. We have added step labels to Figure~\ref{fig:graph_to_circuit_supp} to provide a step-by-step explanation of the scheme.



$ $\\
\textbf{Step-by-step explanation}\\
$ $\\
Step 1. Initial state preparation: The initial state
\begin{align}
	|Init_{N,d}\> = \prod_{j=1}^N (\ha^{j\dagger }_{0,\tilde{0
	}}\ha^{j\dagger }_{0,\tilde{1}}\cdots \ha^{j\dagger}_{0,\widetilde{d-1}})|vac\>
\end{align} can be rewritten in the computational basis $\{0, 1,\cdots, d-1\}$ as 
\begin{align}
	& \prod_{j=1}^N (\ha^{j\dagger }_{0,\tilde{0
	}}\ha^{j\dagger }_{0,\tilde{1}}\cdots \ha^{j\dagger}_{0,\widetilde{d-1}})|vac\> \nn \\
	&= \frac{1}{\sqrt{d}^{Nd}}\prod_{j=1}^N \Big( (\ha^{j\dagger}_{0,0})^d + (-1)^{d-1}(\ha^{j\dagger}_{0,d-1})^d +\sum_{l_j = 1}^{d-2} (-1)^{d-1-l_j}d (\ha^{j\dagger}_{0,0})^{l_j}(\ha^{j\dagger}_{0,d-1})^{d-1-l_j}\ha^{j\dagger}_{0,d-1-l_j} +\eta^j \Big) |vac\> 
\end{align} 
which is obtained with Identities 1, 2 and the relation  $\ha^n\ha^{\dagger m}|vac\> =\frac{m!}{(m-n)!}\ha^{\dagger (m-n)}|vac\>$. In the second line, cases when more than two photons go to modes from 1 to $(d-2)$ 
are all included in $\eta^j$. The reason for grouping them together by $\eta^j$ is because they will be sorted our after the postselection, which we will clarify in Step 6. We extract all the terms that are multiplied by $\eta^j$, hence the above equation is rewritten as 
\begin{align} 
	\frac{1}{\sqrt{d}^{Nd}}\prod_{j=1}^N \Big( (\ha^{j\dagger}_{0,0})^d + (-1)^{d-1}(\ha^{j\dagger}_{0,d-1})^d +\sum_{l_j = 1}^{d-2} (-1)^{d-1-l_j}d (\ha^{j\dagger}_{0,0})^{l_j}(\ha^{j\dagger}_{0,d-1})^{d-1-l_j}\ha^{j\dagger}_{0,d-1-l_j}\Big)|vac\> +\chi_1.
\end{align}   
From here on, $\chi_n$ for each $n$th step will include such terms in each step. 

$ $\\   
Step 2. $CN_d$ operation:
\begin{align}
	\frac{1}{\sqrt{d}^{Nd}}\prod_{j=1}^N \Big( (\ha^{j\dagger}_{0,0})^d +(-1)^{d-1}(\ha^{j\dagger}_{d-1,d-1})^d +\sum_{l_j = 1}^{d-2} (-1)^{d-1-l_j}d (\ha^{j\dagger}_{0,0})^{l_j}(\ha^{j\dagger}_{d-1,d-1})^{d-1-l_j}\ha^{j\dagger}_{d-1-l_j,d-1-l_j} \Big)|vac\> + \chi_2
\end{align}

$ $\\  
Step 3. Rewiring:
\begin{align}
	\frac{1}{\sqrt{d}^{Nd}}\prod_{j=1}^N \Big( (\ha^{j\dagger}_{0,0})^d +(-1)^{d-1}(\ha^{j\oplus_d 1 \dagger}_{d-1,d-1})^d +\sum_{l_j = 1}^{d-2} (-1)^{d-1-l_j}d (\ha^{j\dagger}_{0,0})^{l_j}(\ha^{j\oplus_d 1\dagger}_{d-1,d-1})^{d-1-l_j}\hat{X}^{j\dagger}_{d-1-l_j,d-1-l_j} \Big)|vac\> + \chi_3
\end{align}

$ $\\  
Step 4. $X_d$ and $F_d$: $F_d$ is applied to both wires in each subsystem, hence
\begin{align}
	&\frac{1}{\sqrt{d}^{Nd}}\prod_{j=1}^N \Big( (\ha^{j\dagger}_{0,\tilde{0}})^d +(-1)^{d-1}(\ha^{j\oplus_d 1 \dagger}_{d-1,\tilde{0}})^d +\sum_{l_j = 1}^{d-2} (-1)^{d-1-l_j}d (\ha^{j\dagger}_{0,\tilde{0}})^{l_j}(\ha^{j\oplus_d 1\dagger}_{d-1,\tilde{0}})^{d-1-l_j}\hat{X}^{j\dagger}_{d-1-l_j,d-1-l_j} \Big)|vac\> + \chi_4
\end{align}

$ $\\  
Step 5. $\overline{CN_d}$ operation:
\begin{align}
	\frac{1}{d^{Nd}}\prod_{j=1}^N \Big[
	&\big(\sum_{s=0}^{d-1} \ha^{j\dagger}_{0\ominus_d s,s} \big)^d  +(-1)^{d-1}(\sum_{s=0}^{d-1}\ha^{j\oplus_d 1 \dagger}_{d-1\ominus_d s,s})^d\nn \\
	&+\sum_{l_j = 1}^{d-2} (-1)^{d-1-l_j} d \sqrt{d} \big(\sum_{s=0}^{d-1} \ha^{j\dagger}_{0\ominus_d s,s} \big)^{l_j}\big(\sum_{s=0}^{d-1}\ha^{j\oplus_d 1 \dagger}_{d-1\ominus_d s,s}\big)^{d-1-l_j}\hat{X}^{j\dagger}_{d-1-l_j,d-1-l_j} \Big]|vac\> + \chi_4
\end{align}

$ $\\  
Step 6. Postselection and feed-forward:
From Identities 1 and 2, we can see that when the photons pass through $CN_d$ gates, the only case when one photon goes to $X^j_{d-1-l}$ ($l\in \{1,2,\cdots, d-2\}$) is when $l$ photons goes to $\ha^{\dagger}_{j,0,0}$ and $(d-1-l)$ photons goes to $\ha^{\dagger}_{j,d-1,d-1}$. We use this property to verify surviving terms after the postselection by $D_F$s. 
We can check that when more than two photons go to $X^j_{d-1-l}$ ($l\in \{1,2,\cdots, d-2\}$), such cases are all sorted out by the postselection because $N(d-1)$ photons cannot arrive that $N(d-1)$ $D_F$s for that case. Hence $\chi_4$ vanishes. 
Using  the fact that each subtractor denoted as a dashed box receives photons from only two subsystems, we can see that $\chi_4$ vanishes and the state evolves into

\begin{align}
	\frac{1}{d^{Nd}}\Big( (d!)^N|0,0,\cdots\> +(d!)^N|d-1,d-1,\cdots, d-1\> + (d!\sqrt{d}t)^N\sum_{k=1}^{d-2}|k,k,\cdots, k\>\Big), 
\end{align} which becomes the qudit GHZ state with amplitude carved by BSs with $t=\frac{1}{\sqrt{d}}$. 
The success probability is given from the normalization factor by
\begin{align}
	P_{suc}(N,d) =d\Big(\frac{d!}{d^{d}}\Big)^{2N}. 
\end{align}

\section{Comparison with other schemes for generating the qudit GHZ state}

Here we compare our scheme with previously proposed schemes for generating $|GHZ_N^d\>$. 

\subsection{Deterministic and probabilistic methods}

In optics, we can generate entanglement in two ways--- deterministically or probabilistically. In deterministic methods, entanglement is generated with near-certainty when the process is executed. For example, Ref.~\cite{bell2022protocol} suggested a near-deterministic scheme for generating $|GHZ_N^d\>$ using an array of non-interacting single-photon emitters. Deterministic approaches, in principle, guarantee the generation of entangled states with every attempt, hence having a much higher efficiency compared to probabilistic methods. However, they rely on high-fidelity quantum gates and synchronized single-photon emitters that are technically still out of reach. They are also more susceptible to noise and decoherence in general.

For the case of probabilistic methods, we can again divide them into two types: postselected and heralded schemes. Postselected schemes prearrange which state to
postselect and identifies desired outcomes after conducting detections. Several schemes have been suggested for generating qubit and qudit GHZ states, e.g.,~\cite{erhard2018experimental,chin2021graph,lee2022entangling,bhatti2023generating}. Compared to heralded ones, they generally use smaller resources with simpler circuit elements, which result in higher success probability. For example, Ref.~\cite{bhatti2023generating} used $N$ photons and symmetric multiport splitters to generate qubit $N$-partite GHZ state with success probability $P_{suc} = \frac{N!}{2^{N-1}N^N}$ for odd $N$ and $\frac{N!}{2^{N/2 -1}N^N}$ for even $N$ to the estimated scaling. On the other hand,
since the successful generation of target states must be verified by direct detections, in many quantum tasks they cannot be used as genuine resources.


In contrast, entangled states generated by heralding enable the verification of experimental runs that produce the target states without directly measuring them. Consequently, a heralded  entangled state can become a valuable resource for quantum tasks involving the concatenation of quantum gates. However, it
requires additional particles and modes for the heralding process, making them less efficient and more complex to design. So far as we know, Ref.~\cite{paesani2021scheme} is the only work that proposes linear optical heralded schemes of $|GHZ_N^d\>$ for arbitrary $d$ and $N$. Therefore, it is worth comparing the schemes in the work with ours, which we will execute in the next subsection.

In addition, there also exist schemes for generating $|GHZ_N^d\>$ using other types of particles, e.g., atoms~\cite{cervera2022experimental,zhao2024dissipative} and trapped ions~\cite{upadhyay2024scalable}.  

\subsection{Heralded schemes for generating the qudit GHZ state}

\begin{center}
	\begin{figure}[t]
		\includegraphics[width=1\textwidth]{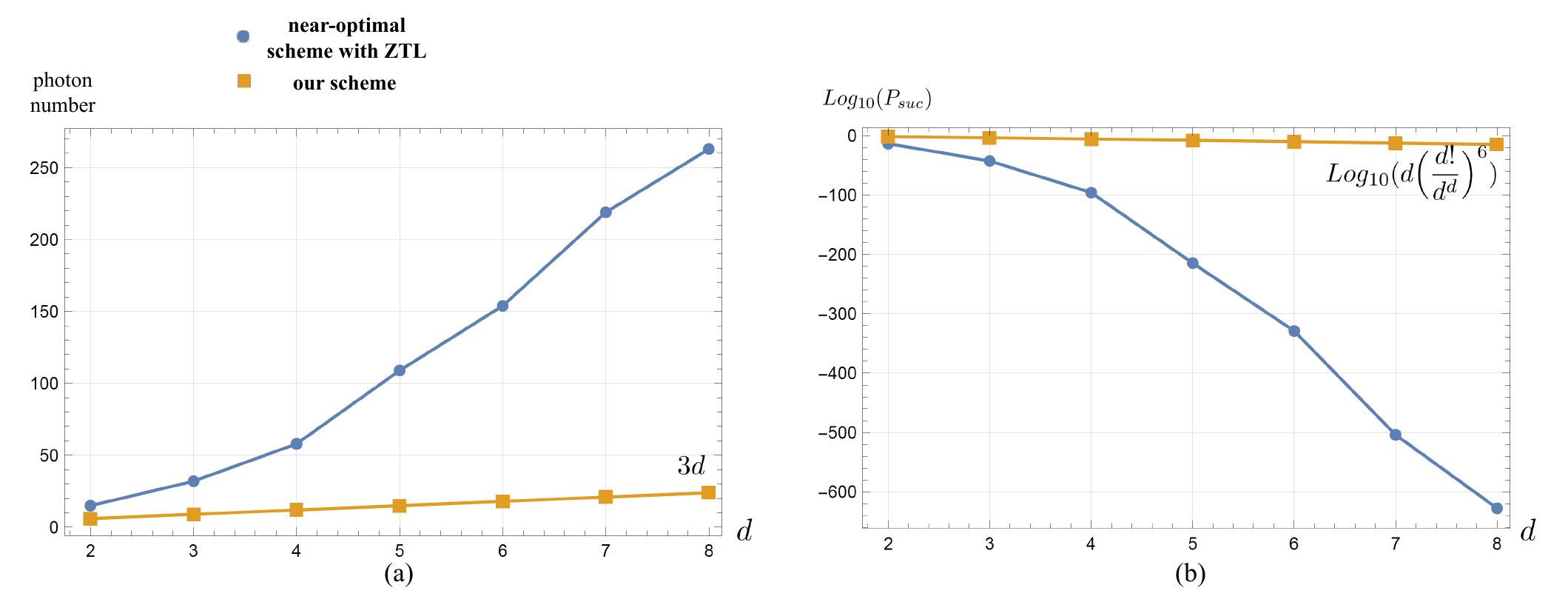}
		\caption{Comparison of our scheme with the near-optimal scheme of Ref.~\cite{paesani2021scheme} for $N=3$ in (a) photon numbers, and (b) the logarithmic scales of the success probabilities. The results in (b) show that the success probability is enhanced exponentially in our scheme.}
		\label{fig:N=3_comparison}
	\end{figure}
\end{center}

Compared to the schemes using the zero-transmission law (ZTL) in Fourier transform interferometers in Ref.~\cite{paesani2021scheme} (which we call `ZTL scheme' from now on), our scheme is substantially more efficient for $N\ge 3$ both in the photon numbers and success probabilities.
For $N=2$ (qudit Bell state), our scheme has lower success probability ($d\Big(\frac{d!}{d^d}\Big)^{4}$ is smaller than $\frac{d(2d-1)!}{(2d+1)^{2d-1}}$ of the ZTL scheme)   but requires less photons (we need $2d$ photons while the ZTL scheme needs $2d+1$).
Since the general ZTL scheme in Ref.~\cite{paesani2021scheme} has a very low efficiency, an alternative near optimal scheme is proposed that can be applied to general $N$ and $d$. 
We can see that the near-optimal solution still needs significantly more photons and  exponentially lower success probability than ours. Figure~\ref{fig:N=3_comparison} compares the required photon numbers and $P_{suc}$ of the near-optimal ZTL scheme with ours for $N=3$ and $2\le d \le 8$.

On the other hand, a brute-fore method is also used in Ref.~\cite{paesani2021scheme} for $N=d=3$ that diminishes the photon number to $25$ with $P_{suc}\sim 10^{-10}$, which is still much less efficient than our scheme that requires $9$ photons with $P_{suc}\sim 3.6\times 10^{-4}$. The $P_{suc}$ can be increased to $\sim 0.8\times 10^{-4}$  with multiplexing, which however is still lower than a quarter of ours.



\section{$dN$ is the minimal photon number for the heralded generation of $|GHZ\>_N^d$ in linear optics}

\begin{figure}[t]
	\centering 
	\includegraphics[width=.35\textwidth]{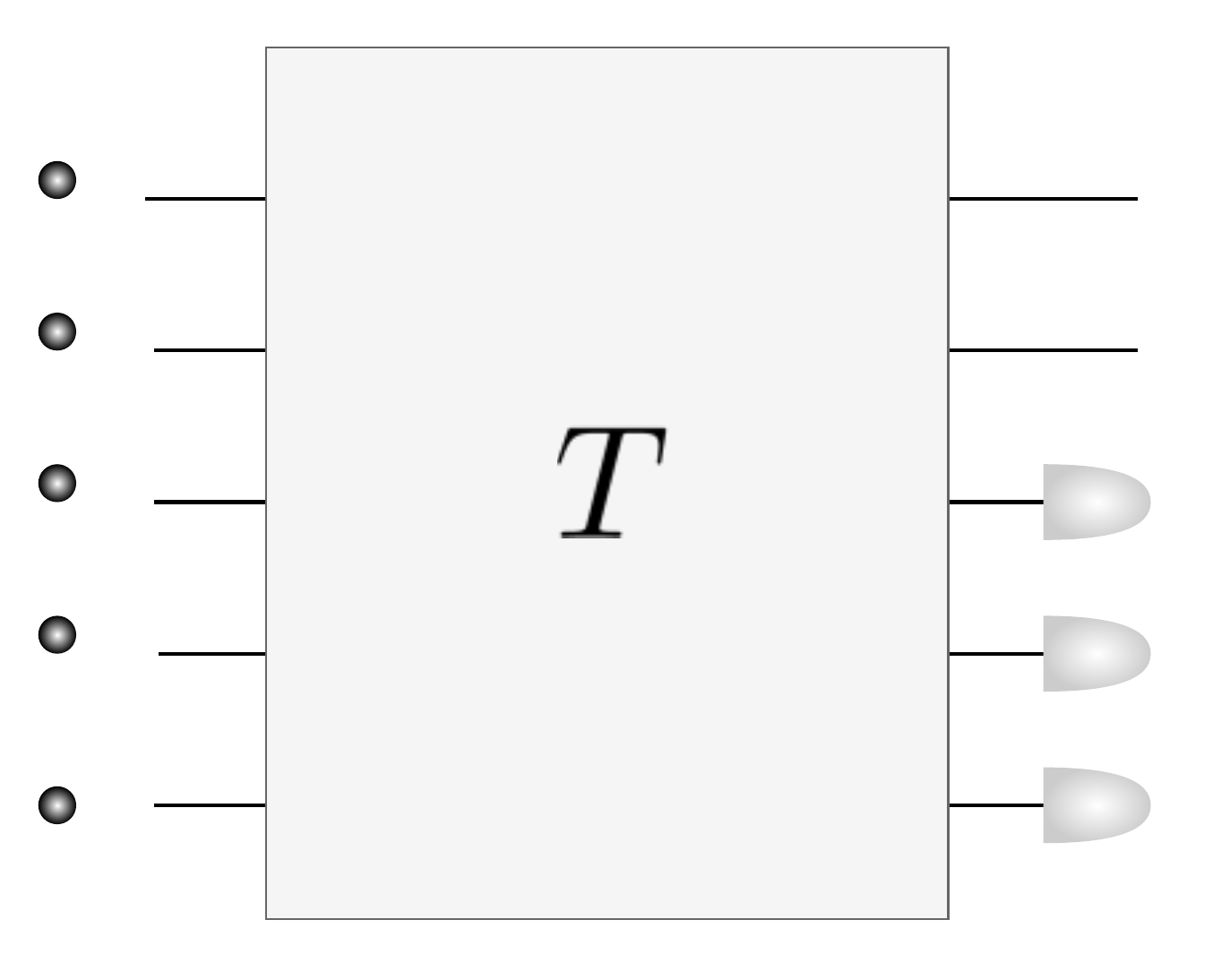}
	\caption{ }
	\label{fig_minimal}
\end{figure}

It is known that the heralded generation of the 2-level Bell state in linear optics requires at least 4 photons~\cite{stanisic2017generating,gubarev2020improved} under the assumption that one photon corresponds to one qubit.
For the most general case, setting that one photon corresponds to one qudit, we can show that \emph{$dN$ is the minimal number of photons for the $d$-level $N$-partite GHZ state}. This implies that \emph{our scheme requires the minimal number of photons to generate the state.}
In this section, we present the proof for arbitrary $d$ and $N$. More exactly, we show that the generalized qudit GHZ state
\begin{align}\label{gen_GHZ}
	\sum_{s=0}^{d-1}x_{s}|\underbrace{s,s,\cdots,s}_{N}\>~~~(x_s \in \mathbb{C},~\sum_{s}|x_s|^2 =1)
\end{align}
cannot be generated in linear optics with less than $(dN-1)$ photons. 

\paragraph*{$(N=2,d=3)$ case.---} As proof of concept, we first explain the simplest nontrivial case, i.e., $(N=2,d=3)$. The mathematical approach for the case will be directly used for the general $d$ and $N$ case in the next paragraph.

Let us consider a linear optical network with 5 photons injected into the input modes and 3 detectors at the output modes (Figure~\ref{fig_minimal}). The state of photon from $j$th input mode ($\had_{j,s_j}$) is transformed under $T$ into $\sum_{j}T^p_j\had_{j r^p_j}$, where $T^p_j$ ($p,j \in \{1,2,3,4,5\}$) is the probability amplitude of the photon from the $p$th input mode to the $j$th output mode ($T_j^p\in \mathbb{C}$,  $\sum_{j}|T^p_{j}|^2=1$) and $|r^p_j\>$ is the internal state of the photon along the path, which we can rewrite as  $|r^p_j\> = \tilde{\a}^p_j|0\> + \tilde{\b}^p_j|1\>+\tilde{\gamma}^p_j|2\>$ ($|\tilde{\a}^p_j|^2+ |\tilde{\b}^p_j|^2+|\tilde{\gamma}^p_j|^2 =1$)\footnote{It is explained in Ref.~\cite{chin2021graph} that this is the simplified expression for the most general linear transformation.}.

Let us assume  that the heralding of the target state corresponds to the case when each detector observes one photon with internal state $|0\>$ without loss of generality. Then the following equality must hold for the heralded scheme to generate the generalized qutrit Bell state:
\begin{align}\label{gen_bell}
	&\sum_{\substack{j,k,l,m,n\\=1}}^5\<vac|_3\<vac|_4\<vac|_5 \ha_{30}\ha_{40}\ha_{50} (T^1_j\had_{j r^1_j})(T^2_k\had_{k r^1_k}) (T^3_l\had_{l r^1_l}) (T^4_m\had_{m r^1_m}) (T^5_n\had_{n r^1_n}) |vac\>_1|vac\>_2|vac\>_3|vac\>_4|vac\>_5 \nn \\
	&= x|0\>_1|0\>_2 +  y|1\>_1|1\>_2+ z|2\>_1|2\>_2.  
\end{align} where where $|vac\>_n$ ($n=\{1,2,3,4,5\}$) is the ground state of the $n$th mode and $x,y,z$ are nonzero complex numbers that satisfy $|x|^2 + |y|^2+|z|^2=1$. The qutrit basis is fixed to a specific computational basis $\{0,1,2\}$ because the basis change can be achieved by local unitary operations.

By the identity $[\ha_{j,\psi},\had_{k,\phi}] = \d_{jk}\<\psi|\phi\>$, the first line of the above equation is rewritten as
\begin{align}\label{gen_bell_2}
	&\frac{1}{2}\sum_{p\neq q\neq r\neq s\neq t=1}^{5}
	T^p_3 T^q_4 T^r_5 \tilde{\a}^p_3\tilde{\a}^q_4\tilde{\a}^r_5 \sum_{a,b=1}^2 T^s_a T^t_b \had_{a r^s_a}\had_{b r^t_b}|vac\>_1|vac\>_2 \nn \\
	&\equiv 
	\frac{1}{2}\sum_{\substack{p,q,r,s,t\\=1} }^5 \chi_{pqrst} X^{pqr} \sum_{a,b=1}^2 T^s_a T^t_b \had_{a r^s_a}\had_{b r^t_b}|vac\>_1|vac\>_2, 
\end{align}
where $\chi_{p_1p_2p_3p_4p_5}\equiv \prod_{a<b}(1-\d_{p_ap_b})$ and  $ X^{pqr} \equiv  T^p_3 T^q_4 T^r_5 \tilde{\a}^p_3\tilde{\a}^q_4\tilde{\a}^r_5 $. Defining $\tilde{X}_{st} \equiv \frac{1}{3!}\sum_{pqr} \chi_{pqrst} X^{pqr}$, Eq.~\eqref{gen_bell} is simplified as
\begin{align}\label{N_5_fin}
	\frac{1}{2}  \sum_{s,t=1}^5 \tilde{X}_{st}\sum_{a,b=1}^2 (T^s_aT^t_b\had_{a r^s_a}\had_{b r^t_b})|vac\>_1|vac\>_2 =   x|0\>_1|0\>_2 +  y|1\>_1|1\>_2+ z|2\>_1|2\>_2.
\end{align}
By setting $(T^s_j\tilde{\a}^s_j = \a^s_j,$ $ T^s_j\tilde{\b}^s_j = \b^s_j,$ $ T^s_j\tilde{\gamma}^s_j = \gamma^s_j )$, the above equation results in the following restrictions:
\begin{align}\label{restriction0}
	& \sum_{s, t=1}^5\tilde{X}_{st}\a^s_1\a^t_1 = \sum_{s, t=1}^5\tilde{X}_{st}\b^s_1\b^t_1=\sum_{s, t=1}^5\tilde{X}_{st}\gamma^s_1\gamma^t_1=\sum_{s, t=1}^5\tilde{X}_{st}\a^s_2\a^t_2=\sum_{s, t=1}^5\tilde{X}_{st}\b^s_2\b^t_2 = \sum_{s, t=1}^5\tilde{X}_{st}\gamma^s_2\gamma^t_2 \nn \\
	&=\sum_{s, t=1}^5 \tilde{X}_{st}\a^s_1\b^t_1 =\sum_{s, t=1}^5\tilde{X}_{st}\a^s_1\gamma^t_1 = \sum_{s, t=1}^5\tilde{X}_{st}\b^s_1\gamma^t_1 =\sum_{s, t=1}^5\tilde{X}_{st}\a^s_2\b^t_2 = \sum_{s, t=1}^5\tilde{X}_{st}\a^s_2\gamma^t_2 = \sum_{s, t=1}^5\tilde{X}_{st}\b^s_2\gamma^t_2  \nn \\
	&= \sum_{s, t=1}^5\tilde{X}_{st}\a^s_1\b^t_2 =\sum_{s, t=1}^5\tilde{X}_{st}\a^s_1\gamma^t_2 = \sum_{s, t=1}^5\tilde{X}_{st}\b^s_1\gamma^t_2 =\sum_{s, t=1}^5\tilde{X}_{st}\a^s_2\b^t_1 = \sum_{s, t=1}^5\tilde{X}_{st}\a^s_2\gamma^t_1 = \sum_{s, t=1}^5\tilde{X}_{st}\b^s_2\gamma^t_1 =0,
\end{align} 
\begin{align}\label{restriction1}
	\sum_{s, t=1}^5\tilde{X}_{st}\a^s_1\a^t_2 = x,~~\sum_{s, t=1}^5\tilde{X}_{st}\b^s_1\b^t_2=y,~~\sum_{s, t=1}^5\tilde{X}_{st}\gamma^s_1\gamma^t_2=z.
\end{align}

We now consider $(\a^p_1, \a^p_2, \b^p_1, \b^p_2, \gamma^p_1, \gamma^p_2)$ as elements of six 5-dimensional complex vectors $(\vec{\a}_1, \vec{\a}_2, \vec{\b}_1,\vec{\b}_2,\vec{\gamma}_1,\vec{\gamma}_2)$. Then, we can see that they must satisfy the following two properties:  

\begin{enumerate}
	\item These six vectors are all nonzero, which is clear from~\eqref{restriction1}. 
	\item  They are not parallel to each other. Indeed, if $\vec{\a}_{1} \newparallel \vec{\a}_{2}$, then $\sum_{s, t=1}^5\tilde{X}_{st}\a^s_1\a^t_1 \propto x\neq 0$ from the first equation of \eqref{restriction1}, which is inconsistent with the first equation of \eqref{restriction0}, etc.   
\end{enumerate}

Using the above properties, we can show that there is no set of six 5-dimensional complex vectors that satisfy the restrictions~\eqref{restriction0} and \eqref{restriction1}. Note that 
$\sum_{s, t=1}^5\tilde{X}_{st}\a^s_1\a^t_1 = \sum_{s, t=1}^5\tilde{X}_{st}\a^s_1\b^t_1=\sum_{s, t=1}^5\tilde{X}_{st}\a^s_1\gamma^t_1=\sum_{s, t=1}^5\tilde{X}_{st}\a^s_1\b^t_2=\sum_{s, t=1}^5\tilde{X}_{st}\a^s_1\gamma^t_2 = 0$ gives that $(\tilde{X}_s)^* \equiv  (\tilde{X}_{st}\a^t_1)^*$ constructs a vector that is orthogonal to $(\vec{\a}_1, \vec{\b}_1,\vec{\b}_2,\vec{\gamma}_1,\vec{\gamma}_2)$. Given that the vectors are 5-dimensional, we see that the five vectors must be linearly dependent. Using the fact that they are not parallel, we can express $\vec{\a}_1$ with a linear combination of the other four vectors, i.e., $\vec{\a}_1 = a\vec{\b}_1+b\vec{\b}_2+c\vec{\gamma}_1+d\vec{\gamma}_2$ ($a,b,c,d \in \mathbb{C}$). Then, by inserting this relation into the first equation of \eqref{restriction1}, we see that 
\begin{align}
	\sum_{s, t=1}^5\tilde{X}_{st}(a\b^s_1+b\b^s_2+c\gamma^s_1+d\gamma^s_2)\a^t_{2} =x    
\end{align} holds, which is inconsistent with the restrictions of \eqref{restriction0}. 
Therefore, there exist no 5-dimensional vectors $(\vec{\a}_1, \vec{\a}_2, \vec{\b}_1,\vec{\b}_2,\vec{\gamma}_1,\vec{\gamma}_2)$ that generate the generalized qutrit Bell state, i.e., the qutrit Bell state cannot be generated in linear optics with 5 photons.  

\begin{figure}[t]
	\centering 
	\includegraphics[width=.35\textwidth]{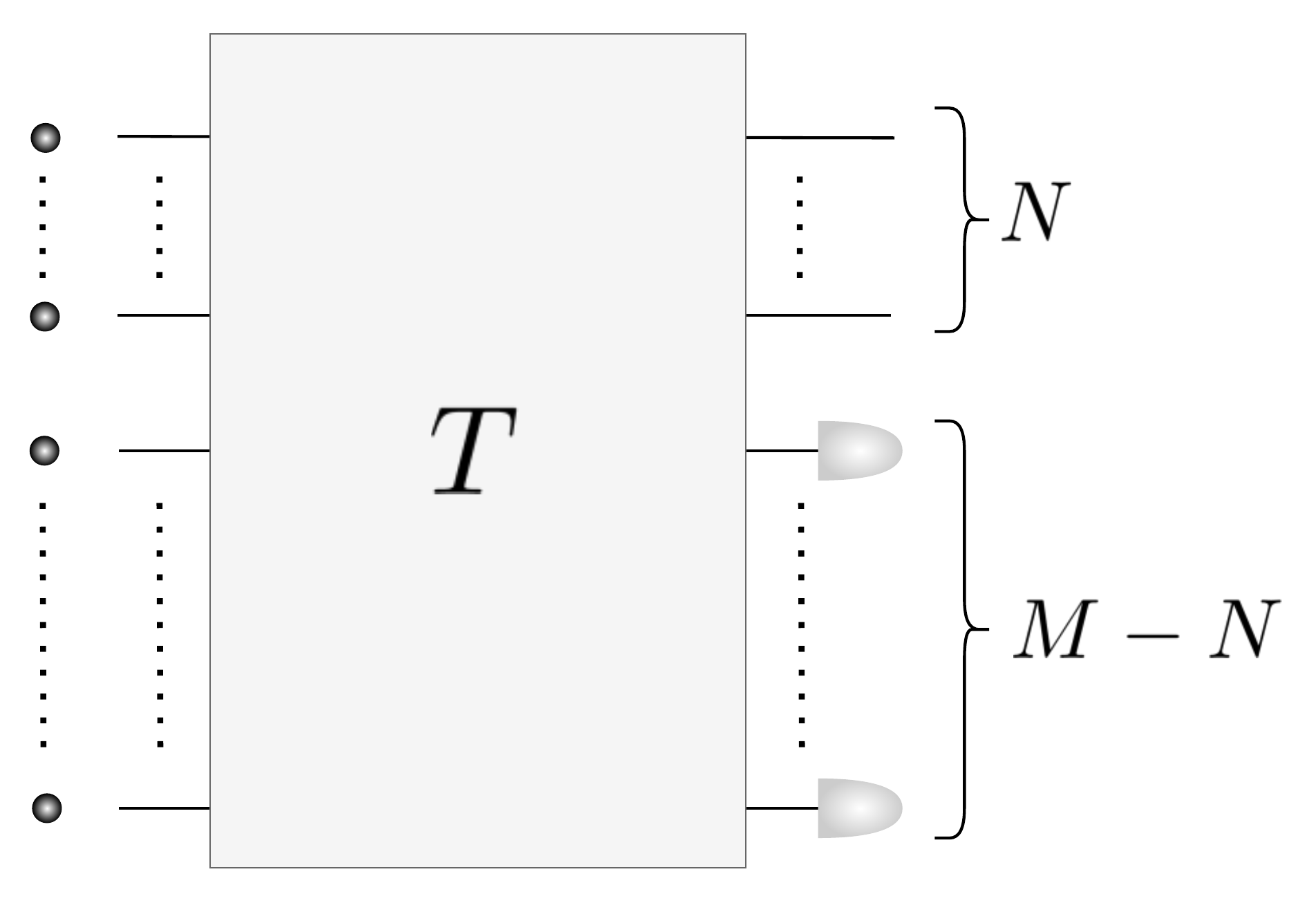}
	\caption{ }
	\label{fig_minimal_gen}
\end{figure}

\paragraph{General $(N,d)$ case.---} The above method can be generalized to the case with arbitrary $d$ and $N$. Now suppose that we have $M$ photons in $M$ modes for generating the generalized qudit $N$-partite GHZ state~\eqref{gen_GHZ} ($N < M$, Figure~\ref{fig_minimal_gen}). Here we will show that the lower limit of $M$ that can generate the state is $dN$.

Now the internal state $|r^p_j\>$ is rewritten in the most general form as  $|r^p_j\> = \sum_{s=0}^{d-1}\tilde{\a}^{ps}_j|s\>$ ($\sum_{s}|\tilde{\a}^{ps}_j|^2=1$).
Then the generalization of Eqs.~\eqref{gen_bell}  and \eqref{N_5_fin} becomes
\begin{align}\label{Nd_gen}
	&\sum_{\substack{j_1,\cdots, j_M=1}}^M\<vac|_{N+1}\cdots \<vac|_{M} \Big[\ha_{N+1,0}\cdots\ha_{M,0} (T_{j_1}^1\had_{j_1 r^1_{j_1}}) (T^2_{j_2}\had_{j_2 r^2_{j_2}}) \cdots (T^M_{j_M}\had_{j_M r^M_{j_M}}) \Big] |vac\>_1\cdots |vac\>_M \nn \\
	&=   \sum_{\substack{p_1,\cdots , p_N\\=1} }^M \tilde{X}_{p_1,\cdots, p_N} \sum_{\substack{j_1,\cdots,j_N\\=1}}^N(\sum_{s_1}\a^{p_1 s_1}_{j_1}\had_{j_1s_1})\cdots (\sum_{s_N}\a^{p_N s_N}_{j_N}\had_{j_Ns_N})|vac\>_1\cdots |vac\>_N\nn \\ 
	&=  \sum_{s=0}^{d-1}x_{s}|\underbrace{s,s,\cdots,s}_{N}\>~~~(x_s \in \mathbb{C},~\sum_{s}|x_s|^2 =1)
\end{align}
where $\tilde{X}_{p_1,\cdots, p_N}  \equiv \sum_{p_{N+1},\cdots ,p_{M}} \chi_{p_1p_2\cdots p_M} \a_{N+1}^{p_{N+1}}\cdots \a_{M}^{p_{M}}$ ($\chi_{p_1p_2\cdots p_N}\equiv \prod_{a<b}(1-\d_{p_ap_b})$) and $\a_{j}^{ps} \equiv T_{j}^{p}\tilde{\a}_{j}^{ps}$.
Eq.~\eqref{Nd_gen} gives the following restriction:

If $s_1=s_2=\cdots =s_N =s$ and $(j_1,j_2,\cdots,j_N)$ is the permutation of  $(1,2,\cdots, N)$, then
\begin{align}\label{restriction_gen_1}
	\sum_{p_1,\cdots, p_N} \tilde{X}_{p_1,\cdots, p_N} \a^{p_1 s}_{1} \a^{p_2 s}_{2}\cdots \a^{p_N s}_{N} = x_{s}.     
\end{align}
Otherwise, 
\begin{align}\label{restriction_gen_2}
	&\sum_{p_1,\cdots, p_N} \tilde{X}_{p_1,\cdots, p_N} \a^{p_1 s_1}_{j_1} \a^{p_2 s_2}_{j_2}\cdots \a^{p_N s_N}_{j_N} = 0.    
\end{align}

We again consider $\a_{j_a}^{p_a s_a}$ as elements of an $M$-dimensional vector $\vec{\a}_{j_a}^{s_a}$. As $j_a\in\{1,2,\cdots, N\}$ and $s_a \in \{0,1,\cdots, d-1\}$, we have $dN$ $M$-dimensional vectors.  These vectors satisfy the following two properties as the  $(N=2,d=3)$ case:

\begin{enumerate}
	\item  They are all nonzero, which is manifest from Eq.~\eqref{restriction_gen_1}.
	\item They are not parallel to each other. Indeed, suppose that there exist $\vec{\a}_{j}^{s}$ and $\vec{\a}_{k}^{t}$ ($j\neq k$ and/or $s\neq t$) that are parallel to each other. Then, from Eq.~\eqref{restriction_gen_1}, 
	\begin{align}
		\sum_{p_1,\cdots, p_N} \tilde{X}_{p_1,\cdots, p_N} \a^{p_1 s}_{1}\cdots  \a^{p_j s}_{j}\cdots \a^{p_N s}_{N} \propto \sum_{p_1,\cdots, p_N} \tilde{X}_{p_1,\cdots, p_N} \a^{p_1 s}_{1}\cdots  \a^{p_j t}_{k}\cdots \a^{p_N s}_{N}  \propto  x_{s}      
	\end{align} holds. However, the second equality of the above equation is inconsistent with Eq.~\eqref{restriction_gen_2}. Therefore, all the vectors are not parallel to each other.
\end{enumerate}

With Eq.~\eqref{restriction_gen_2},  we see that $(\tilde{X}_{p_N}^{s})^* \equiv (\sum_{p_1,\cdots, p_N} \tilde{X}_{p_1,\cdots, p_N} \a^{p_1 s}_{1} \a^{p_2 s}_{j}\cdots \a^{p_{N-1} s}_{N-1})^*$ becomes elements of a vector that is orthogonal to the following $(dN-1)$ vectors:
\begin{align}
	&\vec{\a}_1^{t_1}~(t_1\in\{0,1,\cdots, d-1\}),\nn \\
	&\vec{\a}_2^{t_2}~(t_2\in\{0,1,\cdots, d-1\}),\nn \\
	&\cdots\cdots ,\nn \\
	&\vec{\a}_{N-1}^{t_{N-1}}~(t_{N-1}\in\{0,1,\cdots, d-1\}),\nn \\
	&\vec{\a}_N^{t_N}~(t_N\in\{0,1,\cdots, d-1\}~\textrm{and}~t_N\neq s)
\end{align}
Hence, if $M \le (dN-1)$, the above vectors are linearly dependent. Hence, e.g., $\vec{\a}_1^{0}$ must be a linear function of the other $dN-2$ vectors. This property cannot satisfy Eqs.~\eqref{restriction_gen_1} and \eqref{restriction_gen_2} simultaneously. Hence, $M$ must not be smaller than $dN$ to generate the generalized qudit GHZ state. Since our scheme generates the generalized d-level GHZ state with $dN$ photons, we conclude that $dN$ is the smallest number of photons for generating the generalized d-level GHZ state.

\section{Generating $|GHZ_N^d\>$ in the multi-rail encoding}

Here we present how our scheme (Figure~\ref{fig:graph_to_circuit_supp}) is implemented with the multi-rail encoding for $N=d=3$, which can be directly generalized to any $d$ and $N$.

For $d=3$ case, the multi-rail implementation of linear optical elements are given by
\begin{align}\label{multi_rail_encoding}
	\includegraphics[width=5.5cm]{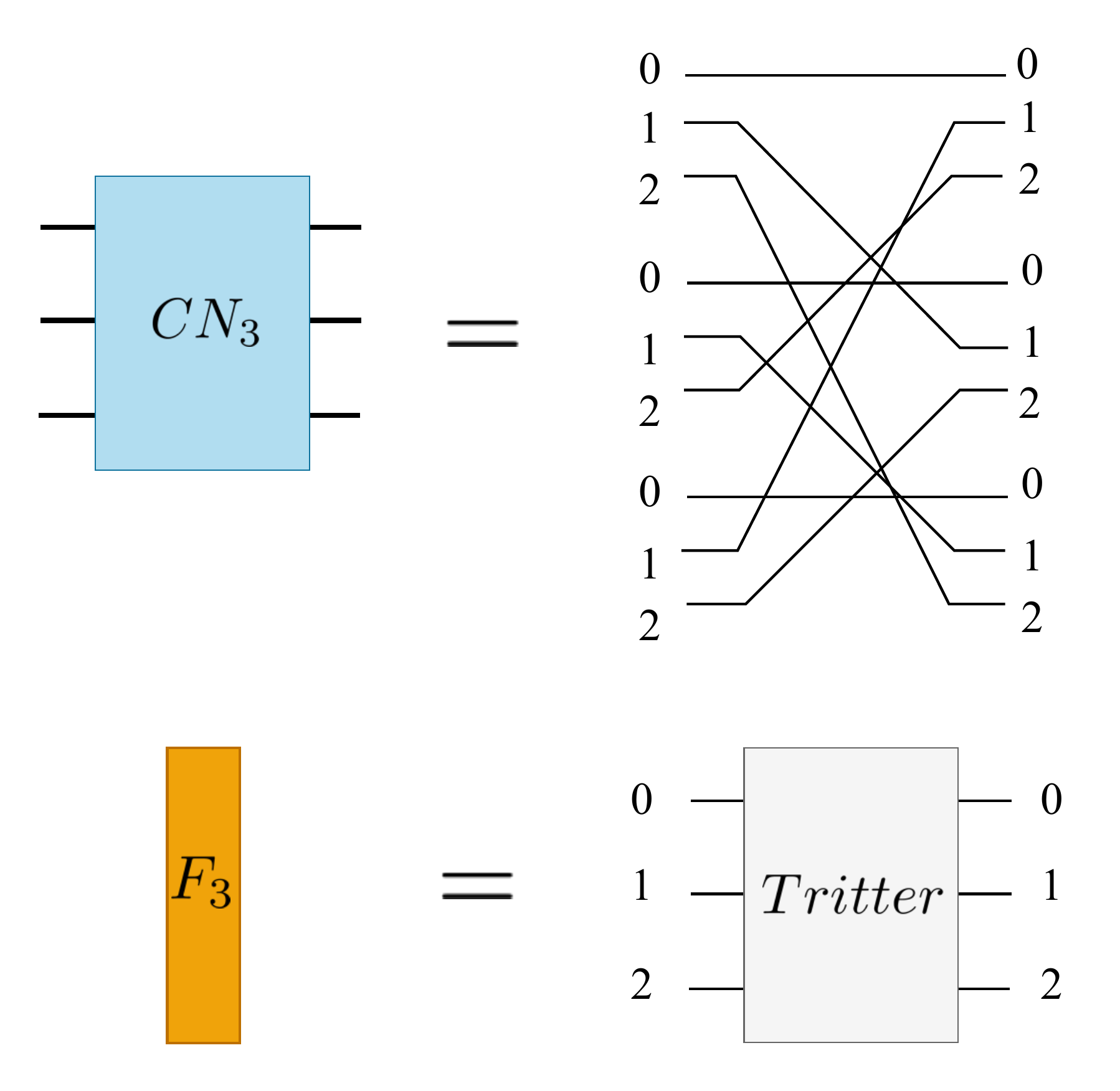},  
\end{align} which can be directly generalized to any $d$-level systems.
Replacing all the linear optical logic gates in Figure~4 of the main content with the operators in~\eqref{multi_rail_encoding}, we obtain a circuit for the $N=3$ qutrit GHZ state in the multirail encoding as follows:
\begin{align}
	\includegraphics[width=.5\textwidth]{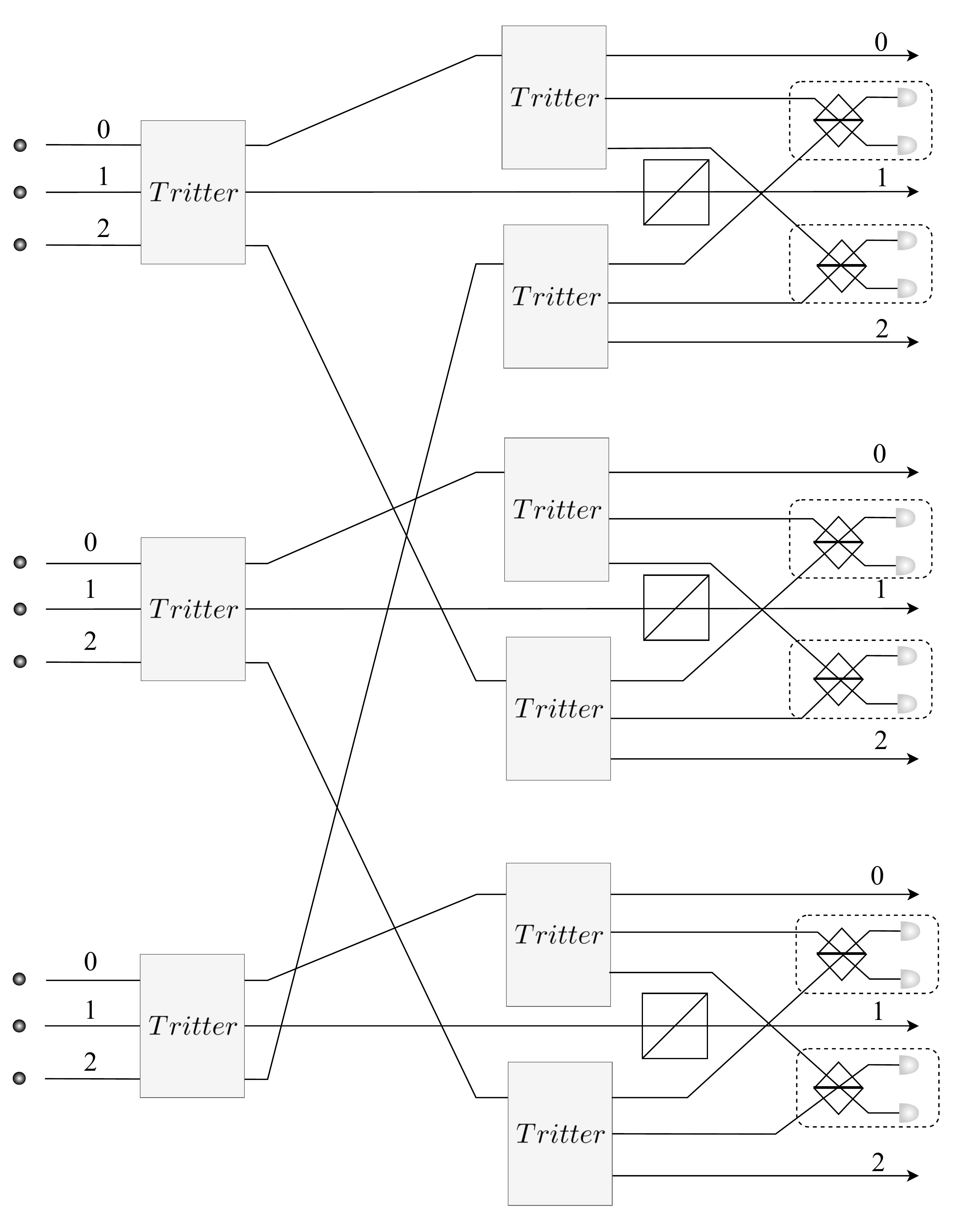} \nn 
\end{align}



\end{document}